\documentclass[amssymb,superscriptaddress,notitlepage,floatfix]{revtex4-1}
\usepackage{amsmath}
\usepackage{amssymb}
\usepackage{amsthm}
\usepackage{amsfonts}
\usepackage{mathrsfs}
\usepackage{listings}
\usepackage{lmodern}
\usepackage{enumerate}
\usepackage{latexsym}
\usepackage{rotating}
\usepackage{float}
\usepackage{psfrag}
\usepackage{bm}
\usepackage{graphicx}
\usepackage[caption=false]{subfig}
\usepackage{blkarray}
\usepackage{array}
\usepackage{color}
\usepackage[normalem]{ulem}
\usepackage{hyperref}

\newcolumntype{L}{>{$}l<{$}}

\newtheorem*{defn*}{Definition}

\renewcommand{\hm}[2]{\mathit{#1}#2}
\newcommand{\bfit}[1]{\mbox{\bfseries{\itshape{#1}}}}
\newcommand{\grel}[1]{\mbox{\textit{\textsf{#1}}}}
\newcommand{\gro}[1]{\mbox{$\mathcal{#1}$}}
\newcommand{\ita}{{\em IT}A }

\begin{document}
\title{Double crystallographic groups and their representations on the Bilbao Crystallographic Server}
\author{Luis Elcoro}
\affiliation{Department of Condensed Matter Physics, University of the Basque Country UPV/EHU, Apartado 644, 48080 Bilbao, Spain}

\author{Barry Bradlyn}
\affiliation{Princeton Center for Theoretical Science, Princeton University, Princeton, New Jersey 08544, USA}
\author{Zhijun Wang}
\affiliation{Department of Physics, Princeton University, Princeton, New Jersey 08544, USA}

\author{Maia~G. Vergniory}
\affiliation{Donostia International Physics Center, P. Manuel de Lardizabal 4, 20018 Donostia-San Sebasti\'{a}n, Spain}
\affiliation{Department of Applied Physics II, University of the Basque Country UPV/EHU, Apartado 644, 48080 Bilbao, Spain}
\affiliation{Max Planck Institute for Solid State Research, Heisenbergstr. 1,
70569 Stuttgart, Germany.}

\author{Jennifer Cano}
\affiliation{Princeton Center for Theoretical Science, Princeton University, Princeton, New Jersey 08544, USA}

\author{Claudia Felser}
\affiliation{Max Planck Institute for Chemical Physics of Solids, 01187 Dresden, Germany}

\author{B. Andrei Bernevig}
\affiliation{Department of Physics, Princeton University, Princeton, New Jersey 08544, USA}
\affiliation{Donostia International Physics Center, P. Manuel de Lardizabal 4, 20018 Donostia-San Sebasti\'{a}n, Spain}
\affiliation{Laboratoire Pierre Aigrain, Ecole Normale Sup\'{e}rieure-PSL Research University, CNRS, Universit\'{e} Pierre et Marie Curie-Sorbonne Universit\'{e}s, Universit\'{e} Paris Diderot-Sorbonne Paris Cit\'{e}, 24 rue Lhomond, 75231 Paris Cedex 05, France}
\affiliation{Sorbonne Universit\'{e}s, UPMC Univ Paris 06, UMR 7589, LPTHE, F-75005, Paris, France}

\author{Danel Orobengoa}
\affiliation{Department of Condensed Matter Physics, University of the Basque Country UPV/EHU, Apartado 644, 48080 Bilbao, Spain}
\author{Gemma de la Flor}
\affiliation{Department of Condensed Matter Physics, University of the Basque Country UPV/EHU, Apartado 644, 48080 Bilbao, Spain}
\author{Mois~I.~Aroyo}
\affiliation{Department of Condensed Matter Physics, University of the Basque Country UPV/EHU, Apartado 644, 48080 Bilbao, Spain}

\begin{abstract}
A new section of databases and programs devoted to double crystallographic groups (point and space groups) has been implemented in the Bilbao Crystallographic Server (http://www.cryst.ehu.es). The double crystallographic groups are required in the study of physical systems whose Hamiltonian includes spin-dependent terms. In the symmetry analysis of such systems, instead of the irreducible representations of the space groups, it is necessary to consider the single- and double-valued irreducible representations of the double space groups. The new section includes databases of symmetry operations ({\tt DGENPOS}) and of irreducible representations of the double (point and space) groups ({\tt REPRESENTATIONS DPG} and {\tt REPRESENTATIONS DSG}). The tool {\tt DCOMPREL} provides compatibility relations between the irreducible representations of double space groups at different $\mathbf{k}$-vectors of the Brillouin zone when there is a group-subgroup relation between the corresponding little groups. The program {\tt DSITESYM} implements the so-called site-symmetry approach, which establishes symmetry relations between localized and extended crystal states, using representations of the double groups. As an application of this approach, the program {\tt BANDREP} calculates the band representations and the elementary band representations induced from any Wyckoff position of any of the 230 double space groups, giving information about the properties of these bands. Recently, the results of {\tt BANDREP} have been extensively applied in the description and the search of topological insulators.
\end{abstract}
\maketitle
\section{Introduction}

The \emph{Bilbao Crystallographic Server} (http://www.cryst.ehu.es) website offers crystallographic databases and programs \cite{aroyo2006}. It can be used free of charge from any computer with a web browser via Internet.  The applications on the server are organized into different sections depending on their degree of complexity, in such a way that the most complex tools make use of the results obtained by the simpler ones. The server is built on a core of databases that includes data from the \emph{International Tables for Crystallography}, Vol. A: \emph{Space- group symmetry} (\onlinecite{ita}, henceforth abbreviated as \ita), Vol. A1: \emph{Symmetry Relations between Space Groups} \cite{ita1} and Vol. E: \emph{Subperiodic groups}  \cite{ite}. A \textbf{k}-vector database with Brillouin-zone figures and classification tables of all the wave vectors for all 230 space groups is also available. Databases of magnetic space groups and magnetic structures have recently been implemented, alongside with a set of computational  tools that facilitate the systematic application of symmetry arguments in the study of magnetic materials \cite{perezmato2015}.  The database of incommensurate structures, hosted by the server, contains both single-modulated structures and composites. Parallel to the databases and the crystallographic software there are a number of programs facilitating the study of specific problems related to solid-state physics, structural chemistry and crystallography involving crystallographic groups, their group-subgroup relations and irreducible representations.

In a number of physical applications it is necessary to include spin-dependent terms in the Hamiltonian of a crystal system: for example, in taking into account relativistic effects in band structure calculations or spin-orbit coupling in crystal-field theory. Then, instead of the crystallographic groups and their representations, it is necessary to consider the so called \emph{double crystallographic groups} and the related \emph{single-valued} or \emph{vector} and \emph{double-valued}  or \emph{spinor} representations. Since their introduction during the first half of the last century \cite{bethe1929,opechowski1940}, the double groups and their representations have been discussed in detail in the literature (see for example, the account given by \onlinecite{bradley1972} and the references therein). Several reference compilations of the character tables of double point groups (\emph{cf.} \onlinecite{altmann1994}), and of the irreducible representations of double space groups \cite{cracknell1979}, abbreviated as CDML exist. However, we are not aware of any online available databases of double crystallographic groups nor of any computing tools for the calculation of their representations. Here, we report on the recently completed development of such programs and their implementation on the Bilbao Crystallographic Server (BCS). 

For what follows, it will be convenient for us to recall briefly some essential features of the double crystallographic groups. Suppose that $\overline{\gro{G}}$ is a point group which consists of pure rotations, \emph{i.e.} a subgroup of the group $SO(3)$. The two-to-one homomorphism $\varphi$ between the group $SU(2)$ of all $(2\times2)$ unitary unimodular matrices onto $SO(3)$ can be used for the formal definition of the double groups $\overline{^d\gro{G}}$ \cite{opechowski1940}: the double group $\overline{^d\gro{G}}$ of a group $\overline{\gro{G}}$ of order $|\overline{\gro{G}}|$ which is a subgroup of the group $SO(3)$, is the abstract group of order 2$|\overline{\gro{G}}|$ having the same multiplication table as the 2$|\overline{\gro{G}}|$ matrices of $SU(2)$ which correspond under $\varphi$ to the elements of the group $\overline{\gro{G}}$. The assignment of the 
$SU(2)$ matrices to the crystallographic symmetry operations used in the databases and programs of the server follows the choice by \onlinecite{altmann1994}.

The kernel of the homomorphism $\varphi:SU(2)\longrightarrow SO(3)$ consists of two elements: $\{1=\begin{pmatrix}1&0\\0&1\end{pmatrix}, {^d1}=\begin{pmatrix}-1&0\\0&-1\end{pmatrix}\}$. As a result, the preimage of each element $R$ of $\overline{\gro{G}}$ consist of two elements of $\overline{^d\gro{G}}$, namely $R$ and $^dR={^d1}R$. 
Formally, one can write
\begin{equation}\label{def_double_group}
\overline{^d\gro{G}}=\{R\}\cup\{^dR\}.
\end{equation}
It is important to note that the subset of elements \{R\} does not form a subgroup of $\overline{^d\gro{G}}$ as it is not closed under the binary group operation. For instance, if $R$ represents a 2-fold rotation or a mirror plane, $R^2=\,^d1$. In that sense it is wrong to refer to eq. (\ref{def_double_group}) as a coset decomposition of $\overline{^d\gro{G}}$ with respect to $\overline{\gro{G}}$. Physically, it is considered that the $SU(2)$ matrices act on the spinors of a 1/2 spin space. Then, the operation $^d1$ is of order 2 and it is often interpreted as a $2\pi$ rotation.

The above statements about double point groups of pure rotations can be generalized in a straightforward way to the case of double point groups that contain improper rotations. For this purpose it is sufficient to indicate that the two $SU(2)$ matrices $\overline{1}=\begin{pmatrix}1&0\\0&1\end{pmatrix}$ and $^d\overline{1}={^d1}\overline{1}=\begin{pmatrix}-1&0\\0&-1\end{pmatrix}$ are assigned to the symmetry operation of inversion.

Consider a space group \gro{G} and its decomposition $\gro{G}:\gro{T}$ into cosets with respect to its translation subgroup \gro{T}, i.e. $\gro{G}=\gro{T}\cup \gro{T}\{R_2|\mathbf{v}_2\} \cup \dots \cup {\gro{T}}\{R_n|\mathbf{v}_n\}$. In a similar way, the symmetry operations of the double space group  $^d\gro{G}$ can be conveniently represented using its coset decomposition with respect to \gro{T} as:
\begin{equation}\label{coset_double}
^d\gro{G}={\gro{T}}\cup {\gro{T}}\{R_2|\mathbf{v}_2\} \cup \dots \cup {\gro{T}}\{R_n|\mathbf{v}_n\}\cup {\gro{T}}\{^d1|\mathbf{o}\}\cup\cdots\cup {\gro{T}}\{{^dR}_n|\mathbf{v}_n\},
\end{equation}
 where $\{R_1|\mathbf{v}_1\}=\{1|\mathbf{o}\}$ is omitted. Here, $R_i$ and ${^dR}_i$ are the elements of the double point group $\overline{^d\gro{G}}$ of $^d\gro{G}$. Consequently, there are two elements $\{R_i|\mathbf{v}_i\}$ and $\{^dR_i|\mathbf{v}_i\}$ of the double space group $^d\gro{G}$, that correspond to every element of the space group $\gro{G}$. The translation subgroup \gro{T} is an invariant subgroup of the double space group $^d\gro{G}$. 
 

In the following, we shall discuss the development and implementation of databases and programs involving the double crystallographic groups for the BCS. We start with the presentation of the double space group database and the retrieval tools that access the stored crystallographic symmetry information (Section \ref{sec:doublegroups}). The introduction to the basic programs available on BCS for the computation of representations of double crystallographic groups is given in Section \ref{sec:representations}. The last sections of the article are devoted to the presentation of the accompanying applications of the representations of double crystallographic groups, such as their compatibility relations (Section \ref{sec:compatibilityrelations}), the site-symmetry approach (Section \ref{sec:sitesymmetry}), and finally, the determination of the band representations and the elementary band representations (Section \ref{sec:bandrepresentations}). In the Appendix we briefly describe the normal-subgroup induction procedure.

\section{\label{sec:doublegroups}Double crystallographic groups}

The double space groups are infinite groups, \emph{i.e.} they contain an infinite number of symmetry operations generated by the set of all translations of the space group. As already noted, a practical way to represent the symmetry operations of the double space group $^d\gro{G}$ is based on the coset decomposition of $^d\gro{G}$  with respect to its translation subgroup $\gro{T}$, \emph{cf.} eq.(\ref{coset_double}). 
The set of coset representatives $\{\{R_i|\mathbf{v}_i\},\{^dR_i|\mathbf{v}_i\}, i=1,\dots, n\}$ of the decomposition $^d\gro{G}:\gro{T}$ (often referred to as \emph{General positions} of $^d\gro{G}$) represents, in a clear and compact way, the infinite number of symmetry operations of the double space group $^d\gro{G}$. The infinite symmetry operations in a coset have the same linear part $R_i$ (or $^dR_i$) while their translation parts differ by lattice translations.
 The translations $\{1|\mathbf{t}\}\in\gro{T}$ form the first coset with the identity $\{1|\mathbf{o}\}$ as a coset representative. The number of cosets is always finite and is equal to the order of the double point group $\overline{^d{\gro{G}}}$ of the double space group $^d\gro{G}$. 

The database of double space groups, available on BCS, includes the lists of the representatives of the general positions of each double space group $^d\gro{G}$. These data can be retrieved using the {\tt DGENPOS} tool \newline(http://www.cryst.ehu.es/cryst/dgenpos), by specifying the sequential \ita  number or the Hermann-Mauguin symbol of the space group \gro{G} corresponding to $^d\gro{G}$. The symmetry operations are specified by their matrix presentations, \emph{i.e} by the $(3\times4)$ matrix-column pairs and the $(2\times2)$ matrices of $SU(2)$, and by Seitz symbols.

 \begin{itemize}
\item Matrix presentations 

With reference to a coordinate system (O, $\mathbf{a}_1,\mathbf{a}_2, \mathbf{a}_3$), consisting of an origin O and a basis  $\mathbf{a}_k$, the symmetry operations of the space group $^d\gro{G}$ are described by matrix-column 
pairs $(\bfit{W},\,\bfit{w})$. The set of translations are represented by the matrix-column pairs (\bfit{I},\ $\bfit{t}_i$), where $\bfit{I}$ is the ($3\times 3$) unit matrix and $\bfit{t}_i$ is the column of coefficients of the translation vector $\mathbf{t}_i$ that belongs to the {\em vector lattice} $\mathbf{L}$ of $^d\gro{G}$. The programs and databases of the double space groups, as well the rest of the computer tools on BCS, use specific settings of space groups (hereafter referred to as standard, or default, settings) that coincide with the conventional space-group descriptions found in \ita. For space groups with more than one description in \ita, the following settings are chosen as standard: \emph{unique axis b} setting, \emph{cell choice 1} for monoclinic groups, \emph{hexagonal axes} setting for rhombohedral groups, and \emph{origin choice 2} (origin on $\overline{1}$ ) for the centrosymmetric groups listed with respect to two origins in \ita.

The pair of symmetry operations $\{\{R_i|\mathbf{v}_i\},\{^dR_i|\mathbf{v}_i\}\}$ related by $\{^d1|\mathbf{o}\}$ have the same matrix-column presentation $(\bfit{W}_i,\,\bfit{w}_i)$ while their $(2\times2)$ matrices differ by a sign. As already noted, the chosen correspondence scheme between the rotations and the $(2\times2)$ matrices follows closely the choice made by \onlinecite{altmann1994}. The shorthand descriptions of the $(3\times4)$ matrix-column pairs and of the $(2\times2)$ matrices of the symmetry operations are also shown in the general-position table. 
\item Seitz symbols

The Seitz symbols $\{R_i|\mathbf{v}_i\}$ of space-group symmetry operations consist of two parts: a rotation (or linear) part $R$ and a translation part $\mathbf{v}$ \cite{glazer2014}. The Seitz symbol is specified between braces and the rotational and the translational parts are separated by a vertical line. The translation parts \textbf{v} correspond exactly to the columns $\bfit{w}$ of the matrix-column presentation $(\bfit{W},\bfit{w})$ of the symmetry operations. The rotation parts $R$ consist of symbols that specify (i) the type and the order of the symmetry operation, and (ii) the orientation of the corresponding symmetry element with respect to the space-group basis. The symbols 1 and $\overline{1}$  are used for the identity and the inversion, $m$ for reflections, the symbols 2, 3, 4 and 6 denote rotations and $\overline{3}$ , $\overline{4}$  and $\overline{6}$ rotoinversions. For rotations and rotoinversions of order higher than 2, a superscript + or - is used to indicate the sense of the rotation. The orientation of the symmetry element is denoted by the direction of the axis for rotations or rotoinversions, or by the direction of the normal to reflection planes. Subscripts of the symbols specify the characteristic direction of the operation: for example, the subscripts $100$, $010$ and $1\overline{1} 0$ refer to the directions $[100]$, $[010]$ and $[1\overline{1} 0]$, respectively.

The symmetry operations of the double space groups are denoted by the \emph{modified} Seitz symbols. The \emph{modified} Seitz symbols include a superscript $d$ added to the rotational part of the symmetry operations  to distinguish between the symmetry operations $\{R|\mathbf{v}\}$ and those obtained by their combinations with  $\{^d1|\mathbf{o}\}$. 
\end{itemize}

\begin{figure}
\begin{center}
\includegraphics[width=0.6\textwidth]{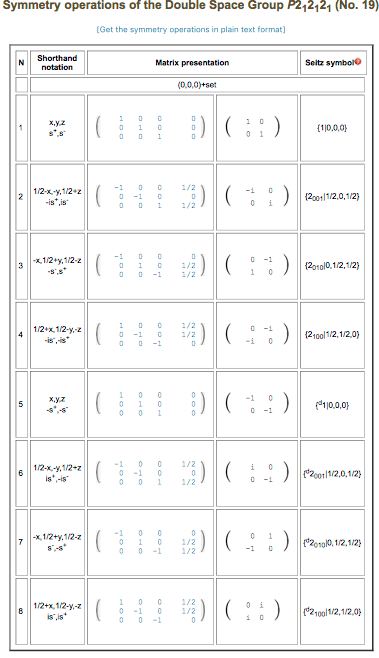}

\caption{Figure 1: Screenshot of the output of the program {\tt DGENPOS}. The figure shows the general-position table of the double space group $\hm{P}{2_12_12_1}$ (No. 19). The symmetry operations are specified by their matrix representations, shorthand notation (coordinate triplets and spin components) and Seitz symbols.}
\label{fig:dgenpos}
\end{center}
\end{figure}

As an example, consider the general-position table of the double space group $\hm{P}{2_12_12_1}$ (No. 19) shown in Fig. \ref{fig:dgenpos}. The listed eight symmetry operations are the chosen coset representatives of the decomposition of $\hm{P}{2_12_12_1}$ with respect to its translation subgroup. The pairs of elements related by $\{^d1|\mathbf{o}\}$ are clearly distinguished by their Seitz symbols $\{R_i|\mathbf{v}_i\}$ and $\{^dR_i|\mathbf{v}_i\}$, \emph{e.g.} $(2) \{2_{001}|1/2,0,1/2\}$ and $(6) \{^d2_{001}|1/2,0,1/2\}$. As explained, the pairs of symmetry operations $\{\{R_i|\mathbf{v}_i\},\ \{^dR_i|\mathbf{v}_i\}\}$
are represented by the same $3\times4$ matrix-column pair while their $2\times2$ matrices differ by a sign.

\section{\label{sec:representations}Representations of the double crystallographic groups}
\subsection{The problem}
There are two programs on the Bilbao Crystallographic Server that compute the irreducible representations (irreps) of space groups explicitly, namely {\tt REPRES} (http://www.cryst.ehu.es/cryst/repres) and {\tt Representations SG}  (http://www.cryst.ehu.es/cryst/ representationsSG). Given a space group $\cal{G}$ and a {\bf k}-vector, both programs calculate the irreps of space groups following the algorithm based on a normal-subgroup induction method, \emph{i.e.} the irreps of a group $\cal{G}$ are calculated starting from those of a normal subgroup $\cal{H}\lhd\cal{G}$. The main steps of the procedure involve the construction of all irreps of $\cal{H}$ and their distribution into orbits under $\cal{G}$, determination of the corresponding little groups and the allowed (small) irreps and finally, construction of the irreps of $\cal{G}$ by induction from the allowed irreps. The labels assigned to the irreps calculated by {\tt REPRES} and {\tt Representations SG} correspond to those used by CDML. The correct assignment of the labels to the irreps calculated by {\tt REPRES}  is achieved with the help of the database of space-group irreps recently developed by \onlinecite{stokes2013}.

Here we report on the development of the program {\tt Representations DSG} (http://www.cryst.ehu.es/cryst/ representationsDSG), available on BCS, for the computation of the irreps of double crystallographic groups. The method for calculating of the irreps of double crystallographic groups is based on a generalization of the normal-subgroup induction procedure for the calculation of space-group irreps implemented in the programs {\tt REPRES} and {\tt Representations SG}.  

To make the exposition self consistent, we include in the Appendix the main concepts and statements of the normal-subgroup induction procedure (for a detailed presentation, \emph{cf}. \onlinecite{aroyo2006}).

\subsection{\label{irrep_method}The method}
The normal-subgroup induction procedure for the calculation of space-group representations (see the Appendix) can be generalized in a straightforward way for the calculations of the representations of the double space groups (\emph{cf.} Miller \& Love, 1967). Consider the coset decomposition of the double group $^d\gro{G}$ with respect to its translation subgroup \gro{T} (\emph{cf.} eq. (\ref{coset_double})):

\begin{equation}
^d\gro{G}=\bigcup_i\gro{T} \grel{q}_i \cup \gro{T}\,{^d\grel{q}}_i.
\end{equation} 

It will be convenient to rewrite the coset decomposition of $^d\gro{G}$ with respect to the group $^d\gro{T}=\gro{T}\otimes\{\{1|000\},\{^d1|000\}\}$:
\begin{equation}
^d\gro{G}={^d\gro{T}}\cup {^d\gro{T}}\grel{q}_2 \cup \cdots \cup {^d\gro{T}}\grel{q}_m.
\end{equation} 

The group ${^d\gro{T}}$ is a trivial central extension of $\gro{T}$ by the group generated by  $\{^d1|000\}$; it is an abelian group, and it is a normal subgroup of $^d\gro{G}$, \emph{i.e.} $^d\gro{G}\triangleright {^d\gro{T}}$. Each irrep $\mathbf{\Gamma}^\textbf{k}$ of \gro{T} generates two irreps $\mathbf{\Gamma}^{\mathbf{k}}$ and $\overline{\mathbf{\Gamma}}^{\mathbf{k}}$ of ${^d\gro{T}}$:

\begin{eqnarray}
\mathbf{\Gamma}^{\mathbf{k}}(\{1|\mathbf{t}\})=\mathrm{exp}(-i\textbf{k}\cdot\textbf{t})\textrm{,}&\hspace{0.15cm}\mathbf{\Gamma}^{\mathbf{k}}(\{^d1|\mathbf{t}\})=\mathrm{exp}(-i\textbf{k}\cdot\textbf{t}),\\
\overline{\mathbf{\Gamma}}^{\mathbf{k}}(\{1|\mathbf{t}\})=\mathrm{exp}(-i\textbf{k}\cdot\textbf{t})\textrm{,}&\hspace{0.15cm}\overline{\mathbf{\Gamma}}^{\mathbf{k}}(\{^d1|\mathbf{t}\})=-\mathrm{exp}(-i\textbf{k}\cdot\textbf{t}).
\end{eqnarray}

The irrep $\mathbf{\Gamma}^{\mathbf{k}}$ is known as a \emph{single-valued} irrep: the same (one-dimensional) matrix represents the elements $\{1|\mathbf{t}\}$ and $\{^d1|\mathbf{t}\}$. On the contrary, the (one-dimensional) matrices of the elements $\{1|\mathbf{t}\}$ and $\{{^d1}|\mathbf{t}\}$ differ by a sign in the \emph{double-valued} irrep $\overline{\mathbf{\Gamma}}^{\mathbf{k}}$.

Note that the term single-valued (double-valued) representation is kept for any representation $\mathbf{D}^{\mathbf{k}} (\overline{\mathbf{D}}^{\mathbf{k}})$ induced from $\mathbf{\Gamma}^{\mathbf{k}} (\overline{\mathbf{\Gamma}}^{\mathbf{k}})$ as all such representations are characterized by the same relationship between the matrices representing the elements related by the operation ${^d\grel{1}}$, namely:

 \begin{eqnarray}
\mathbf{D}^{\mathbf{k}}(\{R|\mathbf{t}\})=\mathbf{D}^{\mathbf{k}}(\{^dR|\mathbf{t}\}),\\
\overline{\mathbf{D}}^{\mathbf{k}}(\{R|\mathbf{t}\})=-\overline{\mathbf{D}}^{\mathbf{k}}(\{^dR|\mathbf{t}\}).
\end{eqnarray}

As the wave vector is left invariant under the action of ${^d\grel{1}}$, the double little cogroup $^d\overline{\gro{G}}\,^{\mathbf{k}}$ of the wave vector $\mathbf{k}$ consists of the sets of elements of $\{R^{\mathbf{k}}\}$and $\{^dR^{\mathbf{k}}\}$ that correspond to the elements of $\overline{\gro{G}}\,^{\mathbf{k}}$, \emph{cf.} eq. (\ref{def_little_cogroup}):
\begin{equation}\label{def_double_little_cogroup}
^d\overline{\gro{G}}\,^{\mathbf{k}}=\{R\}\cup\{^dR\}.
\end{equation}
The group $^d\overline{\gro{G}}\,^{\mathbf{k}}$ is a subgroup of the double point group $^d\overline{\gro{G}}$ of the space group $^d\gro{G}$. The index $\mid{^d\overline{\gro{G}}}/{^d\overline{\gro{G}}}\,^{\mathbf{k}}\mid$ equals the index $\overline{\gro{G}}\,^{\mathbf{k}}$ in $\overline{\gro{G}}$, and for the coset decomposition of $^d\overline{\gro{G}}$ with respect to $^d\overline{\gro{G}}\,^{\mathbf{k}}$ one can choose the same set of coset representatives as of the decomposition of $\overline{\gro{G}}$ with respect to $\overline{\gro{G}}\,^{\mathbf{k}}$. As a result, the star $*\mathbf{k}$ of the wave vector $\mathbf{k}$ in $^d{\gro{G}}$ coincides with $*\mathbf{k}$ in \gro{G}. 

In analogy to the relationship between the little co-group $\overline{\gro{G}}\,^{\mathbf{k}}$ and the little group ${\gro{G}}\,^{\mathbf{k}}$ (\emph{cf.} eq. (\ref{litgr})), the little group $^d\gro{G}\,^{\mathbf{k}}$ can be defined as: 

\begin{equation}
^d\gro{G}^{\mathbf{k}}=\{\{R^{\mathbf{k}}|\mathbf{v}^{\mathbf{k}}\} \in 
\gro{G}|R^{\mathbf{k}} \in \,^d\overline{\gro{G}}\,^{\mathbf{k}}\}.\label{litgr_double} 
\end{equation} 


The allowed single-valued $\mathbf{D}^{\mathbf{k},i}$ and double-valued irreps $\overline{\mathbf{D}}^{\mathbf{k},j}$ of $^d\gro{G}\,^{\mathbf{k}}$ can be determined from the allowed single-valued $\mathbf{\Gamma}^{\mathbf{k}}$ and double-valued irreps $\overline{\mathbf{\Gamma}}^{\mathbf{k}}$ of $^d\gro{T}$ by stepwise induction along the composition series of $^d\gro{G}\,^{\mathbf{k}}$:

 \begin{equation}\label{comp_little_double}
^d\gro{G}^{\mathbf{k}}\rhd{^d\gro{H}}_1^{\mathbf{k}}\rhd\ldots\rhd{^d\gro{H}}_{m-1}^{\mathbf{k}}
\rhd{^d\gro{H}}_{m}^{\mathbf{k}} \rhd\ldots\rhd{^d\gro{H}}_n^{\mathbf{k}}   ={^d\gro{T}}
\end{equation}

where $\mid{^d\gro{H}}_{m-1}^{\mathbf{k}}/{^d\gro{H}}_{m}^{\mathbf{k}}\mid=2\mbox{ or }3$.

The full single-valued irreps $\mathbf{D}^{\mathbf{*k},i}$ and full double-valued irreps $\overline{\mathbf{D}}^{\mathbf{*k},j}$ of the double space group $^d\gro{G}$ are induced from the allowed single-valued and double-valued irreps of the little group $^d\gro{G}^{\mathbf{k}}$. The full irreps are of dimension $r\times s$ where $s$ is the number of arms of the star $*\mathbf{k}$ and $r$ is the dimension of the corresponding allowed irrep of $^d\gro{G}^{\mathbf{k}}$. 
The matrices of the full irreps have a block structure with $s\times s$ blocks, each of dimension $r\times r$.

\subsection{\label{subsec:complexconjugation}Complex conjugation}

For applications involving time-reversal symmetry, it is necessary to recall briefly the classification of irreps with respect to complex conjugation or, as it is commonly referred to, with respect to their \emph{reality} (for more details, \emph{cf.} \onlinecite{bradley1972}). Note that the concepts and results formulated in the following for ordinary representations of groups, are equally valid for single- and double-valued representations of double groups.

An irrep $\mathbf{D}$ of the group $\gro{G}$ is of:
\begin{itemize}
\item[(i)] the \emph{first kind} (or \emph{real}) if $\mathbf{D}$ is equivalent to a group of real matrices;
\item[(ii)] the \emph{second kind} (or \emph{pseudoreal}) if $\mathbf{D}$ is equivalent to $\mathbf{D}^*$ but not to any group of real matrices;
\item[(iii)] the \emph{third kind} (or \emph{complex}) if $\mathbf{D}$ is not equivalent to $\mathbf{D}^*$.
\end{itemize} 

One can show that \cite{herring1937}
\begin{equation}
\frac{1}{\left| \cal{G} \right|}\sum_{j=1}^{|\gro{G}|}\chi\left(g_j^2\right)=\left\{
\begin{array}{rl}
1&\textrm{iff $\mathbf{D}$ is of the first kind},\\
-1&\textrm{iff $\mathbf{D}$ is of the second kind},\\
0&\textrm{iff $\mathbf{D}$ is of the third kind},\\
\end{array}\right.
\label{test_general}
\end{equation}
 where $\chi$ is the character of the irrep $\mathbf{D}$ of \gro{G}.
The obvious difficulties in the direct application of the \emph{reality} test (eq. \ref{test_general}) to space-group irreps can be overcome if  the sum over all operations of \gro{G} can be replaced by a sum over a relatively small number of space-group elements. This can be achieved using the relationship between the space-group irreps and the allowed irreps of the little group from which they are induced. As a result, the reality test for space group irreps can be written in the following form (see \onlinecite{bradley1972} for details of the proof):
the irrep $\mathbf{D}^{*\mathbf{k},j}$ of the space group \gro{G} induced from the allowed irrep $\mathbf{D}^{\mathbf{k},j}$ of the little group $\gro{G}^\mathbf{k}$, $\mathbf{D}^{*\mathbf{k},j}=\mathbf{D}^{\mathbf{k},j}\uparrow\gro{G}$, is of the first, second or third kind according to:

 \begin{equation}
\frac{1}{|\mathbf{H}_{\mathbf{k}}|}\sum_{\{R_s|\mathbf{v_s}\}\in \mathbf{H}_{\mathbf{k}}}\chi^{\mathbf{k},j}\left(\{R_s|\mathbf{v_s}\}^2\right)=1, -1, \textrm{or} \ 0,
\end{equation}
where $\chi^{\mathbf{k},j}$ is the character of the allowed irrep $\mathbf{D}^{\mathbf{k},j}$ of the little group $\gro{G}^\mathbf{k}$ and the sum is restricted to the set $\mathbf{H}_{\mathbf{k}}$ of coset representatives $\{R_s|\mathbf{v_s}\}$ of \gro{G} with respect to \gro{T} whose rotation parts transform $\mathbf{k}$ into a vector equivalent to $-\mathbf{k}$. Obviously, the element $\{R_s|\mathbf{v_s}\}^2$ leaves $\mathbf{k}$ invariant and therefore it belongs to $\gro{G}^\mathbf{k}$.

The irreps of the first and of the second kind are also known as \emph{self-conjugate} while the irreps of the third kind  form pairs of conjugated irreps $(\mathbf{D}, \mathbf{D}^*)$ which, in general, may be induced from allowed irreps of wave vectors belonging to different stars.

The concepts of \emph{physically irreducible} representations or \emph{time-reversal invariant} representations used in some physical applications are closely related to the above-discussed \emph{reality} of the representations. Once the reality of a space-group irrep has been determined and, for the complex irreps, the pairs of conjugated irreps have been identified, the \emph{time-reversal (TR)-invariant} irreps (single- and double-valued irreps of the double space groups) can be constructed according to:
\begin{itemize}
\item If the irrep $\mathbf{D}$ is (a) single-valued and real, or (b) double-valued and pseudo-real, it is TR-invariant.
\item If the irrep $\mathbf{D}$ is (a) single-valued and pseudo-real or (b) double-valued and real, the TR-invariant irrep is the direct sum of $\mathbf{D}$ with itself. The dimension of the TR-invariant irrep doubles the dimension of $\mathbf{D}$.  
The label of the TR-invariant irrep consists of two copies of the label of $\mathbf{D}$. 
\item If $\mathbf{D}_1$ and $\mathbf{D}_2$ form a pair of mutually conjugated irreps, the direct sum of both irreps is TR-invariant. The label of the TR-invariant representation is the union of the labels of the two irreps. 
\end{itemize}

\subsection{The program {\tt Representations DSG}\label{progDRepres}}

An algorithm based on the normal-subgroup procedure for the calculation of the irreps of the double space groups (Section \ref{irrep_method}) is implemented in the program {\tt Representations DSG}. 

As \textbf{Input}, the program needs the specification of 
the double space group $^d\gro{G}$ by the sequential {\em IT}A number of \gro{G}.
(As already noted, the programs of BCS use the standard \ita settings for the description of the space groups.) The {\bf k}-vector data can be introduced by choosing the {\bf k}-vector directly from a table provided by the program, where CDML labels are used to designate the symmetry {\bf k}-vector types. The listed {\bf k}-vector coefficients (called \emph{conventional} \textbf{k}-vector coefficients) refer to a basis $\mathbf{a^*},\mathbf{b^*},\mathbf{c^*}$ which is dual to the conventional \emph{IT}A settings of the space groups. 
 
The \textbf{Output} of the program consists of two tables showing the matrices of the little-group irreps and the matrices of the irreps of the space group (\emph{i.e} the full representations). In detail, the program produces:
\begin{itemize}
\item The table of little-group representations (illustrated by the allowed irreps of the point $X:(0,1/2,0)$ of the double space group P$4/ncc$ (No. 130) shown in Fig. \ref{fig:representationsdsg}).

\begin{figure}
\begin{center}
\includegraphics[width=\textwidth]{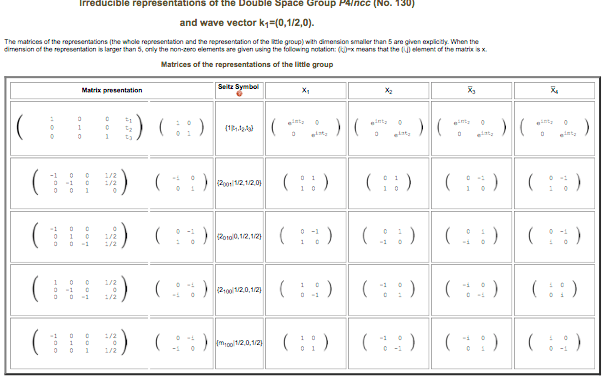}

\caption{Figure 2: Screenshot of the output of the program {\tt Representations DSG}, which shows the allowed irreducible representations of the little group of the point $X\,(0,1/2,0)$ in the double space group P$4/ncc$ (No. 130). Only part of the longer output has been included in the figure.}
\label{fig:representationsdsg}
\end{center}
\end{figure}

The rows of the table are labeled by the symmetry operations of the little group while the columns are specified by the allowed little-group irreps. The symmetry operations are represented by the $(3\times 4)$ matrix-column pairs and $(2\times 2)$ $SU(2)$ matrices, and by the (modified) Seitz symbol. The first row corresponds to a general translation $\{1|\mathbf{t}\}$, while the subsequent rows show the data of the representatives $\{R|\mathbf{v}\}$ of the coset decomposition of the little group with respect to its translation subgroup. (From the fact that any  element of the group $\{R|\mathbf{v}+\mathbf{t}\}$ can be expressed as a combination of a translation $\{1|\mathbf{t}\}$ and a coset representative $\{R|\mathbf{v}\}$ follows that the irrep matrices of $\{R|\mathbf{v}+\mathbf{t}\}$ are equal to products of the type $\mathbf{D}^{\mathbf{k}}(\{R|\mathbf{v}+\mathbf{t}\})=\mathbf{D}^{\mathbf{k}}(\{1|\mathbf{t}\})\mathbf{D}^{\mathbf{k}}(\{R|\mathbf{v}\})$.) The columns of single-valued irreps are followed by those of the double-valued irreps that are ordered according to their dimension. The labels of the single-valued irreps follow the notation of CDML: the labels consist of a $\mathbf{k}$-vector letter(s) and a sequential index. The labels of the double-valued irreps are constructed in a similar way: a bar over the $\mathbf{k}$-vector letter(s) distinguishes the double-valued irreps from the single-valued ones. In the example of the space group P$4/ncc$ (No. 130) and $\mathbf{k}$ vector $X$, $\mathbf{k}=(0,1/2,0)$ (Fig. \ref{fig:representationsdsg}), the single-valued irreps have labels $X_1$ and $X_2$, and the two double-valued irreps are labelled $\overline{X}_3$ and $\overline{X}_4$. 

The irrep matrices are shown explicitly if the dimension of the irrep is $\leq 4$. When the dimension of the irrep is larger than 4, the output shows only the non-zero elements of the matrix, in the format: $(i;j):r$, signifying that the $(i,j)$ element of the matrix has the value $r$.

\item{The full-group irreps (illustrated by the full-group irreps of the double space group P$4/ncc$ at the point $X:(0,1/2,0)$ shown in Fig. \ref{fig:representationsdsg2}).}

\begin{figure}
\begin{center}
\includegraphics[width=\textwidth]{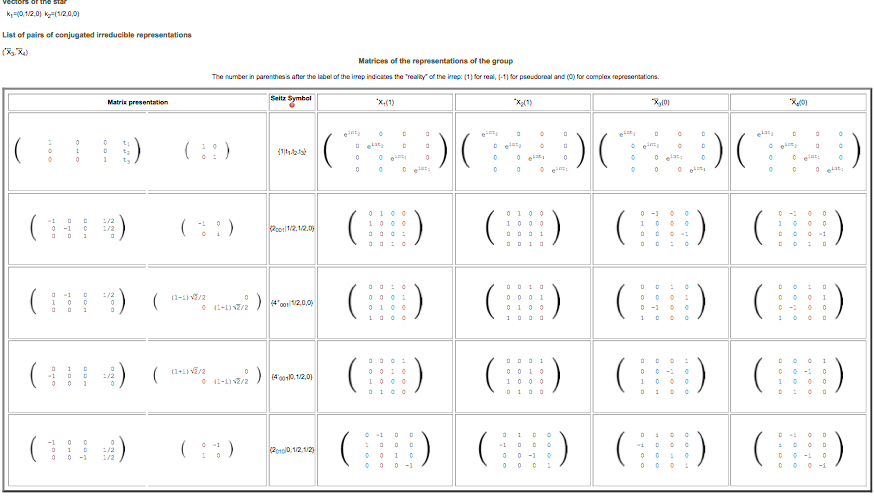}

\caption{Figure 3: Screenshot of the output of the program {\tt Representations DSG}, which shows the irreducible representations of the double space group P$4/ncc$ (No. 130) at the point $X\,(0,1/2,0)$ (full-group). The integer in parenthesis after the irrep label indicates the \emph{reality} of the representation; the corresponding pair of complex-conjugate irreps is also specified. Only part of the longer output has been included in the figure.}
\label{fig:representationsdsg2}
\end{center}
\end{figure}

The indication of the arms of the star $*\mathbf{k}$ precedes the table of the matrices of the full-group irreps $\mathbf{D}^{*k,j}$ of $^d\gro{G}$ induced from the allowed irreps $\mathbf{D}^{\mathbf{k},j}$ of the little group $^d\gro{G}^\mathbf{k}$. The structure and the organization of the data of the full-group irrep table is very similar to that of the little-group irreps: the coset representatives of the decomposition $^d\gro{G}:\gro{T}$ of the double space group with respect to the translation subgroup specify the rows of the table, while the columns correspond to the full space-group irreps. The symmetry operations are described by their matrix-column pairs, $SU(2)$ matrices and Seitz symbols. The labels of the space-group irrep follow the labels of the allowed little-group irreps from which they are induced: \emph{e.g.} as in the case of little-group irreps, a bar over the $\mathbf{k}$ vector letter indicates a double-valued irrep.  

The number in brackets after the irrep label specifies the reality of the irrep per eq. (\ref{test_general}): (1) indicates an irrep of the first kind, {\em i.e.} {\em real}; (-1) - an irrep of the second kind, or {\em pseudoreal};  and (0) - an irrep of the third kind, or {\em complex}. The pairs of the complex conjugated irreps are also indicated in the output. For example, the double-valued irreps  $^*\overline{X}_3$ and $^*\overline{X}_4$ form a complex-conjugated pair as indicated immediately before the table of full irreps (Fig. \ref{fig:representationsdsg2}).\\
As another example, the double-valued irrep $\overline{P}_7$ in the cubic double space group I$a\overline{3}$ (No. 206) is real, so that it doubles when time-reversal is considered. On the contrary, the double-valued irrep $\overline{P}_7$ in the cubic double space group I$4_132$ (No. 214) is pseudoreal and it does not double with time-reversal.
\end{itemize}

\subsection{\label{point_irreps}Representations of the double point groups} 

The crystallographic double point groups and their representations have been extensively discussed in the literature. Sets of irrep compilations can be found, for example, in 
\onlinecite{koster1963}, \onlinecite{bradley1972} and \onlinecite{altmann1994}.  The data on the irreps of the crystallographic double point 
groups are now also online accessible via BCS. The irrep data was obtained by the program {\tt Representations DSG} (\emph{cf.} Section \ref{progDRepres}) for the particular case of $\mathbf{k} = \Gamma (0,0,0)$. 

For each of the 32 crystallographic double point groups, specified by their international (Hermann-Mauguin) 
and Schoenflies symbols, the retrieval tool {\tt Representations DPG} (http://www.cryst.ehu.es/cryst/representationsDPG) displays the following set of tables:

\begin{enumerate}
\item \emph{Character table} 

As usual, the table entries correspond to characters of the irreps (rows) for the conjugacy classes of symmetry operations  (columns) of the chosen double point group. The irreps are labelled according to the notation of \onlinecite{mulliken1933}, by the $\Gamma$ labels given by \onlinecite{koster1963} and by the labels generated by the program {\tt Representations DSG}.  A bar over the irrep label distinguishes the double-valued (known also as {\em spinor}) representations. The distribution into conjugacy classes of the symmetry operations of the double point group (designated by (modified) Seitz symbols), is also indicated.

\item \emph{Table of irrep matrices}

The matrices of the irreps, as calculated by {\tt Representations DSG} for the particular case of $\mathbf{k}=\Gamma\,(0,0,0)$ are also provided by {\tt Representations DPG}. The matrices are explicitly listed for each symmetry operation (rows) and irrep (columns). The symmetry operations of the double point group are specified by the pair of $(3\times 3)$ rotation and $(2\times 2)$ $SU(2)$ matrices, and by the (modified) Seitz symbol. The classification of irreps with respect to complex conjugation is revealed by the number in brackets after the irrep label: as already noted, (1) indicates an irrep of the first kind, {\em i.e.} {\em real}; (-1) - an irrep of the second kind, or or {\em pseudoreal};  and (0) - an irrep of the third kind, or {\em complex}. The pairs of complex conjugated irreps are also listed.
\end{enumerate}

\begin{figure}
\begin{center}
\includegraphics[width=0.85\textwidth]{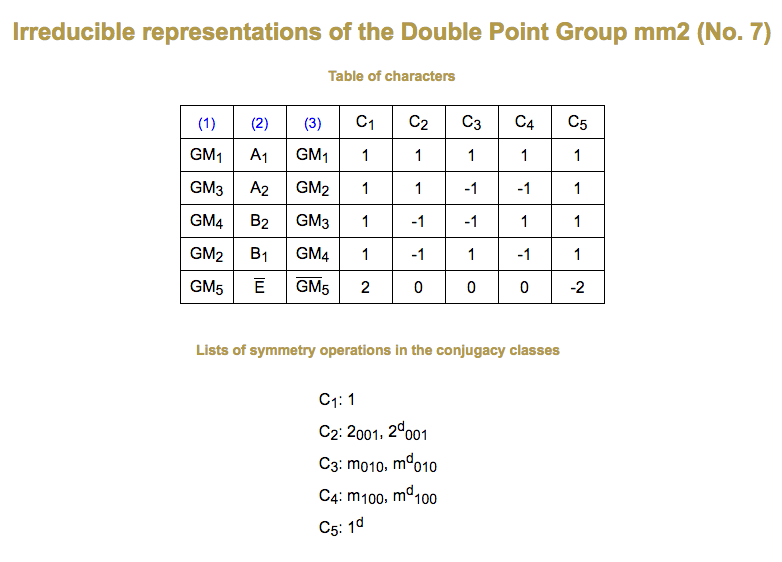}

\caption{Figure 4: Screenshot of the output of the program {\tt Representations DPG} which shows the character table of the double point group $\hm{}{mm2} (C_{2v})$. The irreps are labelled according to the notation of (1)   \onlinecite{koster1963}, (2) \onlinecite{mulliken1933} and (3) by the labels generated by the program. The distribution into conjugacy classes of the symmetry operations is also indicated. Only the first part of the output has been included in the figure.}
\label{fig:character_point}
\end{center}
\end{figure}

\begin{figure}
\begin{center}
\includegraphics[width=0.9\textwidth]{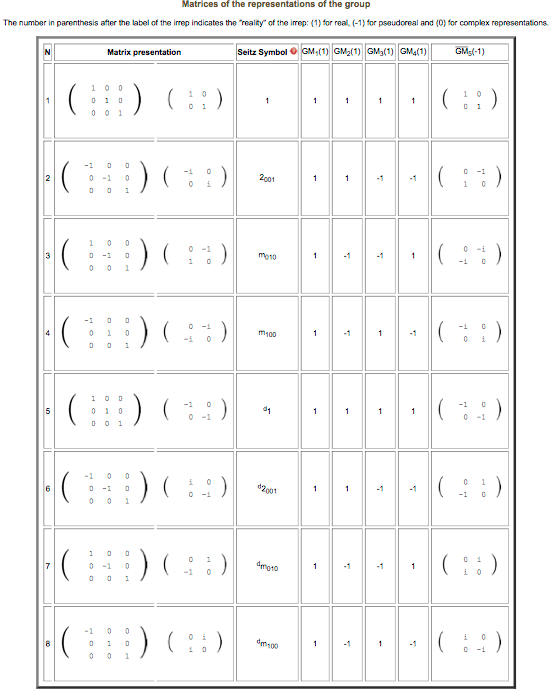}

\caption{Figure 5: Screenshot of the output of the program {\tt Representations DPG} which shows the irrep matrices of the double point group $\hm{}{mm2} (C_{2v})$. The symmetry operations are specified by the matrix presentations and Seitz symbols. The integer in parenthesis after the irrep label indicates the \emph{reality} of the representation. Only the last part of the output has been included in the figure.}
\label{fig:matrices_point}
\end{center}
\end{figure}

Screenshots of the character table and the table of irrep matrices for the double point group $\hm{}{mm2} (C_{2v})$ are shown in Fig. \ref{fig:character_point} and Fig. \ref{fig:matrices_point}.

\section{\label{sec:compatibilityrelations}Compatibility relations}
\subsection{The problem}
The so-called \textit{compatibility relations} have different applications in solid-state physics as, for example, in the analysis of the electronic band structures or phonon dispersion curves.
The compatibility relations are essential in the study of connectivity of energy bands as we move in a continuous way from one \textbf{k}-vector point to a neighbouring one with a different symmetry, or in crystal-field splitting problems that arise when a high-symmetry crystal field is perturbed by a field of lower symmetry.    

From a group-theoretical point of view, the compatibility relations correspond to the so-called \emph{subduction} relationships between  the little-group representations of different \textbf{k}-vectors of the same space group $\cal{G}$ whose little groups form a group-subgroup pair. Let $\cal{G}$ and $\cal{H}$ be two groups with group-subgroup relation $\cal{G} > \cal{H}$ with $ n = | \cal{G} | / | \cal{H} |$ being the index of $\cal{H}$ in $\cal{G}$. Consider an \emph{irrep} $\mathbf{D}_{\beta}= \{\mathbf{D}_{\beta}(g), g\in\cal{G} \}$ of a group $\cal{G}$: The subduction of $\mathbf{D}_{\beta}$ to the subgroup $\cal{H}$ results in a representation of the subgroup, the so-called \emph{subduced} representation, formed by the matrices of those elements of $\cal{G}$ that also belong to the subgroup $\cal{H}$, that is, $\mathbf{D}_{\beta}(h)$ with $h\in\cal{H}<\cal{G}$. This subduced representation $(\mathbf{D}_{\beta} \downarrow \cal{H})=\mathbf{D}^S$ is in general reducible, and is decomposable into irreps $\mathbf{d}_{\alpha}$ of $\cal{H}$:

\begin{equation}
(\mathbf{D}_\beta \downarrow {\cal{H}})(h) = \mathbf{D}^S(h) \sim \bigoplus_\alpha n_\alpha^{(\beta)}\mathbf{d}_\alpha(h),\qquad h\in\cal{H}.
\end{equation}

The multiplicities $n_\alpha^{(\beta)}$ of the \emph{irreps} $\mathbf{d}_{\alpha}$ of $\cal{H}$ in the subduced representation $\mathbf{D}_\beta \downarrow \cal{H}$ can be calculated by the \emph{reduction formula} (known also as the Schur orthogonality relation or \emph{`magic' formula}):

\begin{equation}\label{eq:magic}
n_\alpha^{(\beta)} = \frac{1}{|\cal{H}|}\sum_h\chi^S(h)\chi_\alpha(h)^*,
\end{equation}
where $\chi^S(h)$ is the character of the subduced representation and $\chi_\alpha(h)$ is the character of the irrep $\mathbf{d}_\alpha$ for the same element $h \in \cal{H}$.\\

Consider two \textbf{k}-vectors \textbf{k}$_1$ and \textbf{k}$_2$ where \textbf{k}$_2$ = \textbf{k}$_1$ + $\boldsymbol{\kappa}$ with $\boldsymbol{\kappa}$ - an infinitesimal \textbf{k}-vector; \emph{e.g.} \textbf{k}$_1$ could be the wave vector of a symmetry point (line) 'sitting' on a symmetry line (plane) \textbf{k}$_2$. The little group ${\cal{G}}^{\textbf{k}_2}$ is in general a subgroup of the little group ${\cal{G}}^{\textbf{k}_1}$ and the compatibility relations between the \textit{irreps} at \textbf{k}$_1$ and \textbf{k}$_2$ (in the limit  $\kappa \rightarrow 0$) can be established by the subduction of the representations \textbf{D}$^{\textbf{k}_1,i}$ of the little group ${\cal{G}}^{\textbf{k}_1}$ onto the little subgroup ${\cal{G}}^{\textbf{k}_2}$. In the following we say that these two $\mathbf{k}$-vectors are \emph{connected}.

Tables of compatibility relations of double space groups can be found in Miller \& Love (1967) but the compiled tables  provide only the relationships between the irreps of \textbf{k}-vector points and lines; neither the relations between the \textbf{k}-vector lines and planes nor the relations between the \textbf{k}-vector points and planes are made available. The program {\tt DCOMPREL}  (http://www.cryst.ehu.es/cryst/dcomprel) calculates the compatibility relations between the little-group irreps of the double space groups for any pair of symmetry-related wave vectors.  Compatibility relations between little-group irreps of ordinary space groups can be calculated by the program {\tt COMPATIBILITY RELATIONS}, also accessible in BCS (http://www.cryst.ehu.es/rep/rep$\_$correlation$\_$relations).

\subsection {The method}
Essentially, the same algorithm is implemented in the programs {\tt COMPATIBILITY RELATIONS} and {\tt DCOMPREL}: it follows closely the subduction procedure explained above applied to the special case of little-group representations. Consider two symmetry-related wave vectors \textbf{k}$_1$ and \textbf{k}$_2$, where \textbf{k}$_2$ = \textbf{k}$_1$ + $\boldsymbol{\kappa}$ with ${\cal{G}}^{\textbf{k}_2}<{\cal{G}}^{\textbf{k}_1}$. The matrices of an irrep of the little group ${\cal{G}}^{\textbf{k}_1}$ associated to the symmetry operations that belong to ${\cal{G}}^{\textbf{k}_2}$ form a representation of the little group of $\mathbf{k}_2$, which in general, is reducible. The compatibility relations are extracted from the decomposition of the subduced representation ${\cal{G}}^{\textbf{k}_1} \downarrow {\cal{G}}^{\textbf{k}_2}$ into irreps of ${\cal{G}}^{\textbf{k}_2}$:

\begin{equation}
\mathbf{D}^{\textbf{k}_1,i} \downarrow {\cal{G}}^{\textbf{k}_2}\sim \displaystyle\bigoplus_{j=1}^s m_{\textbf{k}_2,j} \mathbf{D}^{\textbf{k}_2,j},
\end{equation}

where $m_{\textbf{k}_2,j}$ represents the multiplicity of the irrep $\mathbf{D}^{\textbf{k}_2,j}$ in the subduced representation. Given the characters $\chi_{\mathbf{D}^{\textbf{k}_1,i}}$ of $\mathbf{D}^{\textbf{k}_1,i}\downarrow {\cal{G}}^{\textbf{k}_2}$ and $\chi_{\mathbf{D}^{\textbf{k}_2,j}}$ of $\mathbf{D}^{\textbf{k}_2,j}$ for all $g\in{\cal{G}}^{\textbf{k}_2}$ (\emph{e.g.} calculated by the program {\tt Representations DSG}), the multiplicities can be obtained by a variation of the reduction formula:

\begin{equation}\label{magic_inf}
m_{\textbf{k}_2,j} = \frac{1}{|{\cal{G}}^{\textbf{k}_2}|}  \displaystyle\sum_{g} \chi_{\mathbf{D}^{\textbf{k}_2,j}}(g) \chi_{\mathbf{D}^{\textbf{k}_1,i}}(g)^*,   \quad \quad \forall g \in {\cal{G}}^{\textbf{k}_2}.
\end{equation}

The expression (\ref{magic_inf}) is not convenient for practical calculations due to the infinite order of the little group ${\cal{G}}^{\textbf{k}_2}$. However, the summation in eq. (\ref{magic_inf}) can be first performed over all translations in ${\cal{G}}^{\textbf{k}_2}$ which will reduce the infinite sum to a finite one over the set $\mathbf{G}^{\textbf{k}_2}_o$ of the representatives $g_k$ of the decomposition of ${\cal{G}}^{\textbf{k}_2}$ in cosets with respect to its translation subgroup:

\begin{equation}
m_{\textbf{k}_2,j} = \frac{1}{n}  \displaystyle\sum_{k=1}^n \chi_{\mathbf{D}^{\textbf{k}_2,j}}(g_k) \chi_{\mathbf{D}^{\textbf{k}_1,i}}(g_k)^*;   \quad \quad \forall g_k \in {\mathbf{G}}^{\textbf{k}_2}_o.
\end{equation}
The number \textit{n} of coset representatives $g_k$ is equal to the order of the little co-group $\overline{\cal{G}}^{\textbf{k}_2}$.

\subsection{The program {\tt DCOMPREL}}
The  procedure described above for the calculation of the compatibility relations between the little-group irreps of space groups is implemented in the program \texttt{DCOMPREL}. The program is available on BCS, and for a number of cases the results have been successfully checked against the compatibility-relation data listed in \onlinecite{miller1967}. 
\paragraph{Input:}
The specification of the double space group by the \ita number leads to a menu with the list of the different symmetry $\mathbf{k}$-vectors for the group chosen. The choice of a $\mathbf{k}$-vector, $\mathbf{k}_1$, produces a second output with the list of all $\mathbf{k}$-vectors  that can be connected to it. The user can then ask for the compatibility relations between the chosen $\mathbf{k}_1$-vector and a single $\mathbf{k}$-vector or between $\mathbf{k}_1$ and all the $\mathbf{k}$-vectors in the list. 
\paragraph{Output:}
After a summary of the input data, a table with the compatibility relations between the little-group \textit{irreps} of the selected \textbf{k}-vectors is shown.

\begin{figure}
\begin{center}
\includegraphics[width=0.55\textwidth]{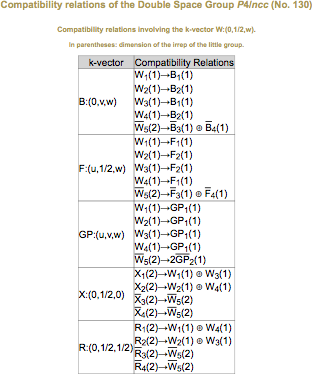}
\caption{Figure 6: Screenshot of the output of the program {\tt DCOMPREL} which shows all the compatibility relations which involve the irreps of the $\mathbf{k}$-vector $W(0,1/2,w)$ in the double space group $\hm{P}{4/ncc}$ (No. 130). The first column gives the list of $\mathbf{k}$-vectors symmetry-related to $W$, and the second column shows the corresponding compatibility relations. In some cases, $W$ is the high-symmetry point of the pair (as in $W\to B,F,GP$) while in others $W$ is the point of lower symmetry (as in $R\to W$). Only part of the longer output has been included in the figure.}
\label{fig:compatibility}
\end{center}
\end{figure}

Fig. \ref{fig:compatibility} shows a partial output of the program {\tt DCOMPREL} of the compatibility relations of the $\mathbf{k}$-vector line $W(0,1/2,w)$ with all symmetry-related wave vectors in the double space group $\hm{P}{4/ncc}$. Note that in some cases, $W$ is the high-symmetry point of the pair (as in $W\to B,F,GP$) while in others $W$ is the point of lower symmetry (as in $R\to W$). The number in parenthesis indicates the dimension of the representation.

\subsection{Example: electronic bands of germanium}
\begin{figure}
\begin{center}
\includegraphics[width=\textwidth]{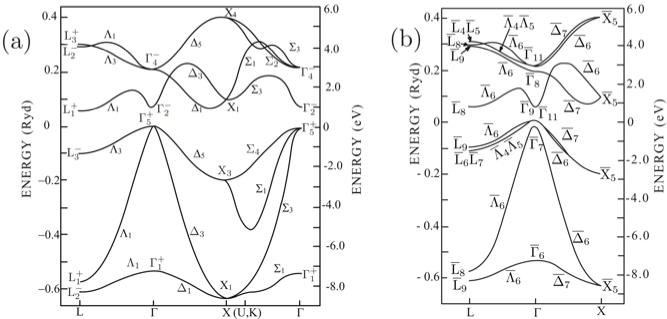}
\caption{Figure 7: Electronic bands of germanium (a) without and (b) with spin-orbit coupling (adapted from Dresselhaus et al. (2008). The labels of the electronic bands correspond to the irreps labels used by {\tt REPRESENTATIONS DSG}.}
\label{fig:dresselhaus}
\end{center}
\end{figure}

\begin{figure}
\begin{center}
\includegraphics[width=0.6\textwidth]{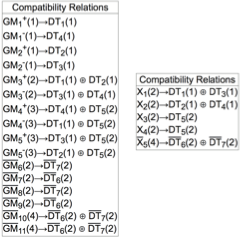}
\caption{Figure 8: Compatibility relations between the point $\Gamma$ and the line $\Delta$ (left) and between the point $X$ and  the line $\Delta$ (right)  of the space group F$d\bar{3}m$, given by the program {\tt DCOMPREL}.}
\label{fig:dresselhauscomp}
\end{center}
\end{figure}

As an application of the use of the representations of the double space groups and the compatibility relations we consider the electronic bands of germanium, with space group F$d\bar{3}m$ (No. 227) and occupied atomic position $8a$ $(1/8,1/8,1/8)$. Figure \ref{fig:dresselhaus} reproduces the figures 12.10 and 14.41 in \onlinecite{dresselhaus2008}. The labels of the irreducible representations have been changed to the labels used in {\tt REPRESENTATIONS DSG}. Figure \ref{fig:dresselhaus} (a) shows the band structure when the spin-orbit coupling is not considered and the relevant irreps are the single-valued irreducible representations, and Figure \ref{fig:dresselhaus} (b) reproduces the band structure when the spin-orbit coupling is considered and the relevant irreps are the double-valued irreducible representations. Figure \ref{fig:dresselhauscomp} shows the compatibility relations between the $\Gamma(0,0,0)$ point and the $\Delta(0,v,0)$ line and between the $X(0,1/2,0)$ point and the $\Delta$ line in the double space group F$d\bar{3}m$.  It can be checked that these compatibility relations agree with the paths $\Gamma \leftrightarrow\Delta\leftrightarrow X$ in Figures \ref{fig:dresselhaus}(a) and \ref{fig:dresselhaus}(b). In some cases, these relations could be useful to identify the irreducible representations at the $\vec{k}$ points of high symmetry. The compatibility relations give the connectivities between the branches of the electronic band along the Brillouin zone and the dimensions of the irreps at each point give the degeneracies of the energy.

\section{\label{sec:sitesymmetry}Site-symmetry approach}
%
\subsection{The problem}

The symmetry of the \emph{extended states} (phonons, soft modes, electronic bands, \emph{etc.}) of crystal structures (described by the irreps of the space group of the crystal) over the entire Brillouin zone is related to the symmetry of \emph{localized states} (local atomic displacements, atomic orbitals, \emph{etc.}) of the constituent structural units (described by the irreps of the local symmetry groups). A procedure for the determination of such a relationship is very useful, as it allows the prediction of the symmetry of the possible extended states starting from the crystal structural data. In particular, it is useful in distinguishing topological materials \cite{NaturePaper}.
Formally, the procedure relating localized and extended crystalline states can be described by induction of a representation of a space group $\cal{G}$ from a finite subgroup $\cal{H}$, followed by a reduction into irreps. In other words, the induction method permits the calculation of the symmetry of the compatible extended states transforming according to irreps of crystal space group $\cal{G}$ induced by a localized state described by an irrep of the local or site symmetry group $\cal{H}=\cal{S}$. However, the calculation of the space-group irreps induced by the irreps of the site-symmetry group is not an easy task: the fact that the site-symmetry group $\cal{S}$ (isomorphic to a point group) is a subgroup of infinite index of $\cal{G}$ implies that the representation of $\cal{G}$ induced by an irrep of $\cal{S}$ must be of infinite dimension, and therefore difficult to calculate directly. The so-called \emph{site-symmetry approach} resolves this problem by applying the \emph{Frobenius reciprocity theorem}, which states that the multiplicities of the irreps of a group $\cal{G}$ in the induced representation from an irrep of a subgroup $\cal{H}$ of $\cal{G}$ can be determined from the multiplicities of the irreps of \gro{H} in the representations subduced from $\cal{G}$ to \gro{H}.
(For a detailed presentation of the method, its discussion and applications, \emph{cf.} \onlinecite{evarestov1997}). In this way, the knowledge of the subduced representation of the irreps of the space group onto the site-symmetry group, which are relatively easy to compute once the space group irreps are known, is enough to obtain all the necessary information about the representations induced by the irreps of the site-symmetry group into the space group.

The site-symmetry method is implemented in two programs of BCS: the program {\tt SITESYM} for ordinary representations of space groups, and the program {\tt DSITESYM} for the double space groups. The following explanations of the site-symmetry method are equally valid for the cases of space groups and double space groups.

\subsection{The method} 
The objective of the site-symmetry programs is to find the symmetry of the crystal extended states induced by localized states of some of the constituent structural units. Examples of the application of this method can be found in the supplementary material of \onlinecite{NaturePaper}. In group-theoretical terms, this task requires the derivation of the irreps of a space group $\cal{G}$ at any point in the reciprocal space (which classify the extended states of the structure) induced by the irreps of the site-symmetry group of a Wyckoff position (according to which localized states are classified). Two basic concepts of representation theory, namely \emph{subduction} and \emph{induction} are essential for the site-symmetry approach. As already explained, the \emph{subduction coefficients} specify the decomposition of a representation of a group $\cal{G}$ into irreps of one of its subgroups $\cal{H}< \cal{G}$ while   
the induction procedure permits the construction of a representation of $\cal{G}$ starting from a representation of $\cal{H}$. Consider the decomposition of $\cal{G}$ in cosets with respect to $\cal{H}$ with coset representatives $\{\grel{q}_m, m=1,\dots,n\}$, \emph{cf.} eq. (\ref{coset}). 
If $\mathbf{d}_\alpha=\{\mathbf{d}_\alpha(h),h\in\cal{H}\}$ is an irrep of $\cal{H} < \cal{G}$, then the matrices of the induced representation $(\mathbf{d}_\alpha \uparrow \cal{G}) = \mathbf{D}^I$ of $\cal{G}$ are constructed as follows:

\begin{equation}\label{ind-1}
\mathbf{D}^I(g)_{kt,js} =\left\{
\begin{array}{l l}
\mathbf{d}_\alpha(q_k^{-1} g q_j)_{ts} & \text{if }  q_k^{-1} g q_j \in \cal{H}\\
\rule{0pt}{4ex}0 & \text{if } q_k^{-1} g q_j \not \in \cal{H}
\end{array}\right.,
\end{equation}
where $k,j = 1,\dots,n$ and $t,s=1,\dots,m$ with $m$ the dimension of the irrep $\mathbf{d}_\alpha$ of $\cal{H}$. Its characters are given by:
\begin{equation}
\chi^I_\gamma(g) =\sum_j \chi^\alpha(q_j^{-1}g q_j),
\end{equation}

where $\chi^\alpha(q_j^{-1}g q_j)$ is the trace of the $j$-th diagonal block of $\mathbf{D}^I(g)$. In general, the induced representations are reducible, and as such it is possible to decompose them into irreps $\mathbf{D}_\beta$ of $\cal{G}$: 
\begin{equation}\label{ind-dec}
(\mathbf{d}_\alpha \uparrow{\cal{G}}) = \mathbf{D}^I \sim \bigoplus_\beta n^{(\alpha)}_\beta \mathbf{D}_\beta,
\end{equation}
where $n^{(\alpha)}_\beta$ are the multiplicities of the irreps of $\cal{G}$ in the induced representation, and it is possible to calculate them applying the reduction formula (\ref{eq:magic}):

\begin{equation}\label{ind-multi}
n^{(\alpha)}_\beta = \frac{1}{|\cal{G}|}\sum_g\chi^I(g)\chi^*_\beta(g),
\end{equation}

where $\chi_\beta(g)$ is the character of the element $g$ of the irrep $\mathbf{D}_\beta$ of $\cal{G}$.

The dimension of the induced representation can be directly read off the equation for its construction, eq.(\ref{ind-1}): 
\begin{equation}\label{eq:ind-dim}
dim(\mathbf{d}_\alpha \uparrow {\cal{G}})= dim (\mathbf{d}_\alpha) \frac{|\cal{G}|}{|\cal{H}|}.
\end{equation} 

This result points to the difficulties for the direct calculation of a representation of a space group $\cal{G}$ induced from an irrep of a finite subgroup $\cal{H}$ of $\cal{G}$: due to the infinite index of $\cal{H}$ in $\cal{G}$, the dimension of the representation of $\cal{G}$ induced from an irrep of $\cal{H}$ is infinite. By means of the site-symmetry approach it is possible to determine the multiplicities $n^{(\alpha)}_\beta$ of an irrep of $\cal{G}$ in the induced representation without the necessity of constructing the infinite-dimensional representation. As stated above, the method is based on the Frobenius reciprocity theorem, according to which the multiplicity of an irreducible irrep $\mathbf{D}_\beta(g)$ of $\cal{G}$ in a representation $\mathbf{d}_\alpha \uparrow\cal{G}$ of $\cal{G}$ induced by an irrep $\mathbf{d}_\alpha$ of $\cal{H}<\cal{G}$ is equal to the multiplicity of the irrep $\mathbf{d}_\alpha$ of $\cal{H}$ in the representation $\mathbf{D}_\beta \downarrow \cal{H}$ subduced by $\mathbf{D}_\beta$ of $\cal{G}$ to $\cal{H}$,  \emph{i.e.}  $n_\alpha^{(\beta)}=n_\beta^{(\alpha)}$. In other words, in order to calculate the frequencies $n_\beta^{(\alpha)}$, eq.(\ref{ind-dec}), it is sufficient to calculate the multiplicities $n_\alpha^{(\beta)}$ of $\mathbf{d}_\alpha$ in the subduced representation $\mathbf{D}_\beta \downarrow \cal{H}\sim\bigoplus_\alpha$ $n_{\alpha}^{(\beta)}d_{\alpha}$.

As already discussed, the irreps of a space group $\cal{G}$ are classified according to the reciprocal space wave vectors and for each vector $\bf{k}$ there is a finite set of irreps of $\cal{G}$. The main idea of the site-symmetry approach is that although the representation induced by an irrep of a site-symmetry group in a space group $\cal{G}$ has infinite dimension, it is possible to know the part of this induced representation that corresponds to any set (finite) of irreps of $\cal{G}$ for a given wave vector $\bf{k}$ (calculated in advance). As in most applications only the irreps of $\cal{G}$ related to a few wave vectors are of interest, this partial knowledge of the induced representation proves to be sufficient.

The algorithmic procedure, based on the site-symmetry approach and implemented in the site-symmetry programs {\tt SITESYM}  and {\tt DSITESYM}, is the following:
\begin{enumerate}
\item Consider an occupied Wyckoff position and determine its site-symmetry group $\cal{S}$. Note that it is not sufficient to determine the point group isomorphic to the site-symmetry group: it is necessary to obtain the set of space-group symmetry operations, with their rotation and translations parts, that belong to the site-symmetry group, since the representation matrices of symmetry operations related by translations are, in general, different.
\item Calculate the irreps of the space group $\cal{G}$ for the wave vectors $\bf{k}$ of interest. The BCS program {\tt REPRES} for ordinary space-group representations (or {\tt Representations DSG} for single- and double-valued irreps of double space groups) provides the irrep matrices for any element of the space group (or of the double space group).
\item From the obtained space-group irreps, calculate the representations subduced to the site-symmetry group $\cal{S}$ obtained in the first step.
\item From the irreps of the site-symmetry group (tabulated by the BCS programs {\tt POINT} \cite{aroyo2006} for ordinary point-group representations or {\tt Representations DPG} for single- and double-valued irreps of double point groups), and making use of the reduction formula (\ref{eq:magic}), calculate the multiplicities of the site-symmetry irreps in the subduced representations.
\item Apply the Frobenius reciprocity theorem to obtain the multiplicities of the irreps in the induced representation of $\cal{G}$ from the multiplicities of irreps of $\cal{S}$.
\end{enumerate}

\subsection{The programs {\tt SITESYM} and {\tt DSITESYM}}
This algorithm for the calculation of the multiplicities of space-group irreps in the representations induced by the irreps of a site-symmetry group ($\cal{S}<\cal{G}$) has been implemented in the site-symmetry programs of BCS. The description that follows refers to the program {\tt DSITESYM} (http://www.cryst.ehu.es/cryst/dsitesym) but similar explanations are also applicable to the program {\tt SITESYM} (http://www.cryst.ehu.es/cryst/sitesym). 

The site symmetry approach allows the determination of the symmetry relationships between the extended (Bloch) and localised (Wannier-type) electronic states in crystals. As an example of the utility of {\tt DSITESYM} we will consider the determination of the symmetry of electronic states in (GaAs)$_m$(AlAs)$_n$ semiconductor superlattices grown along the [001] direction, a problem discussed in detail in \onlinecite{kitaev1997}. The output of the program will be illustrated by the specific calculations of the symmetry relationships  between the electronic band states at the point $\vec{k}=M(1/2,1/2,0)$ of the double space group P$\overline{4}m2$ (No. 115) (the structure-symmetry group of (GaAs)$_m$(AlAs)$_n$) and the localised states of atomic orbitals at the Wyckoff position $2g$ $(0,1/2,z)$. In other words, the multiplicities of the irreps of the double space group P$\overline{4}m2$ at the point $\vec{k}=M(1/2,1/2,0)$ shown in the output of {\tt DSITESYM}, describe the transformation properties of the extended electronic states induced by the irreps of the double site-symmetry group of the Wyckoff position $2g$ $(0,1/2,z)$ according to which the localised one-electron wave functions transform.

\paragraph{Input:}
The required input data include the specification of the double space group, of the occupied Wyckoff position and of the wave-vector label of the space-group irreps whose induced-representation multiplicities are to be calculated. The information is entered using three forms: in the first one, the double space group is specified by its \ita sequential number; the wave vector of interest and the occupied Wyckoff position can be selected from the corresponding lists produced by the program.

\paragraph{Output:}
After a header that reproduces the input data, the program displays the following tables with results of the intermediate steps of the procedure (the induction table, that lists the final results of the site-symmetry calculations, is at the bottom of the  screen (v)):
\begin{enumerate}
\item[(i)] \emph{List of operations of the double site-symmetry group $\cal{S}$}: each of the symmetry operations of the double space group that leaves the Wyckoff position representative point invariant, is represented in a $x,y,z$ and in a matrix notation. Labels ($g_1,...,g_n$) necessary for later referencing are assigned to each element of $\cal{S}$.

The double site-symmetry group of the position 2$g \ (0,\frac12,z)$ of the double space group $\hm{P}{\bar 4m2}$ is 
formed by 8 symmetry operations as shown in the screenshot (Fig. \ref{fig:symmetryoperations}) of the program {\tt DSITESYM}.

\begin{figure}
\begin{center}
\includegraphics[width=0.7\textwidth]{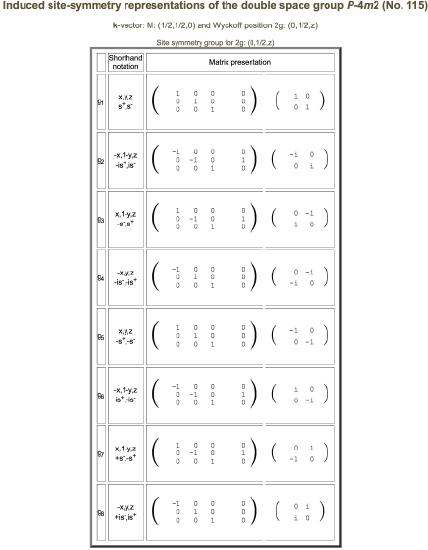}
\caption{Figure 9: Screenshot of the partial output of the program {\tt DSITESYM} which shows the 8 symmetry operations of the double site-symmetry group of the Wyckoff position 2$g \ (0,\frac12,z)$ of the double space group $\hm{P}{\bar 4m2}$ (No. 115). The symmetry operations are specified by the matrix presentations and shorthand notation.}
\label{fig:symmetryoperations}
\end{center}
\end{figure}

\item[(ii)] \emph{Character table of the double point group}. The table reproduces the character table of the irreps of the double point group isomorphic to the site-symmetry group, as provided by the program {\tt Representations DPG} of BCS. The labels of the irreps are given in notations by \onlinecite{mulliken1933} and as generated by the program. 
The site symmetry group of $(0,\frac12,z)$ is isomorphic to the double point group $mm2$ and has five irreps. Its character table is shown in the Fig.  \ref{fig:charactertable}.
\begin{figure}
\begin{center}
\includegraphics[width=\textwidth]{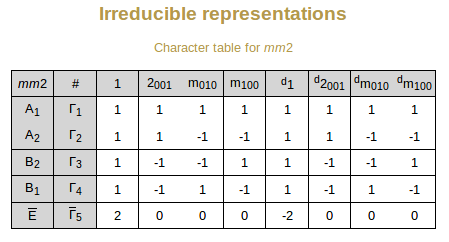}
\caption{Figure 10: Screenshot of the partial output of the program {\tt DSITESYM} which shows the chatacter table of the double point group $mm2$, isomorphic to the site-symmetry group of the Wyckoff position $(0,1/2,z)$ of the space group $\hm{P}{\bar 4m2}$ (No. 115). The irreps are labelled according to the notation of (1) \onlinecite{mulliken1933} and (2) the labels generated by the program.}
\label{fig:charactertable}
\end{center}
\end{figure}

\item[(iii)] \emph{Table of characters of the subduced representations.} The program {\tt Representations DSG} calculates the characters of the elements of the site-symmetry group \gro{S} (obtained in the first step) for each of the irreps $\mathbf{D}_\beta$ of ${^d\cal{G}}$ of the selected wave vector. In this way, we obtain the characters of the subduced representations $\mathbf{D}_\beta \downarrow \cal{S}$ of $\cal{S}$.

The double space group $\hm{P}{\bar 4m2}$ has seven irreps for the wave-vector $M$ which subduce seven representations of the double site-symmetry group $\cal{S}$=$mm2$ with the characters shown in Fig. \ref{fig:characterssubducedirreps}.
\begin{figure}
\begin{center}
\includegraphics[width=0.75\textwidth]{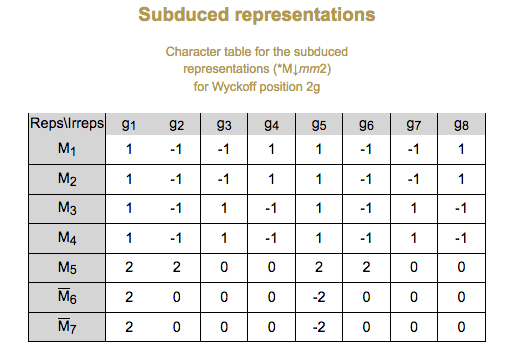}
\caption{Figure 11: Screenshot of the partial output of the program {\tt DSITESYM} which shows the characters of the subduced representations of the site-symmetry group $\cal{S}$=$mm2$ by the irreps of the double space group $\hm{P}{\bar 4m2}$ (No. 115) at the $\mathbf{k}$-point $M$. The columns are specified by the symmetry operations of the site-symmetry group, \emph{cf.} Fig.\ref{fig:symmetryoperations}. }
\label{fig:characterssubducedirreps}
\end{center}
\end{figure}

\item[(iv)] \emph{Table of the decompositions of the subduced representations}. The multiplicities $n_\alpha^{(\beta)}$ of the double point-group irreps $\mathbf{d}_{\alpha}$ of $\cal{S}$ in the subduced representations $\mathbf{D}_{\beta} \downarrow \cal{S}$ are obtained by the application of the reduction formula, \emph{cf.} eq. (\ref{eq:magic}). In the example, the decompositions of the representations $^{*}M_i \downarrow mm2$, $i=1,\ldots,7$, into irreps of $mm2$ are shown in Fig. \ref{fig:decomposition}:
\begin{figure}
\begin{center}
\includegraphics[width=0.55\textwidth]{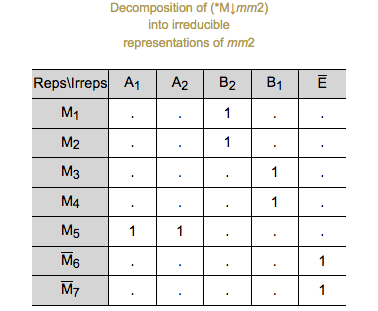}
\caption{Figure 12: Screenshot of the partial output of the program {\tt DSITESYM} which shows the decompositions of the subduced representations $M_i \downarrow mm2$ (see Fig. \ref{fig:characterssubducedirreps}) into irreps of $mm2$, \emph{cf.} Fig. \ref{fig:charactertable}.}
\label{fig:decomposition}
\end{center}
\end{figure}

\item[(v)] \emph{Table of induced representations.}
According to Frobenius reciprocity theorem, the multiplicities $n_\beta^{(\alpha)}$ of the irreps of ${^d\cal{G}}$ for a given \textbf{k}-vector in the representations $\mathbf{d}_\alpha \uparrow{^d\cal{G}}$ (induced from the irreps $\mathbf{d}_\alpha$ of the site-symmetry group $\cal{S}$) are obtained by "transposing" the table of the decompositions of the subduced representations $\mathbf{D}_{\beta} \downarrow \cal{S}$.

The table of representations of the double space group $\hm{P}{\bar 4m2}$ at the point $M$ induced by the irreps of the site-symmetry group $mm2$ of the Wyckoff position 2$g$ is shown in Fig. \ref{fig:representationsbar4m2}. The rows of the table correspond to the irreps $\mathbf{d}_\alpha$ of the site-symmetry group $mm2$ (cf. Fig. \ref{fig:charactertable}); the entries in each row indicate the multiplicities of the $M$-irreps of  $\hm{P}{\bar 4m2}$ in the (infinite) induced representation $\mathbf{d}_\alpha\uparrow \hm{P}{\bar 4m2}$: 
\begin{figure}
\begin{center}
\includegraphics[width=0.55\textwidth]{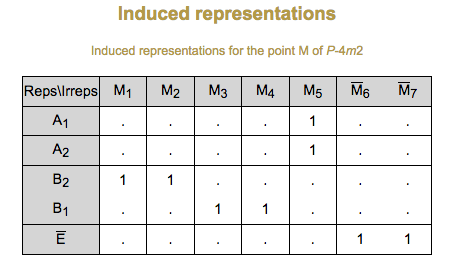}
\caption{Figure 13: Screenshot of the partial output of the program {\tt DSITESYM} which shows the representations of the double space group $\hm{P}{\bar 4m2}$ (No. 115) at the point $M$ induced by the irreps of the site-symmetry group $mm2$ of the Wyckoff position 2$g$.}
\label{fig:representationsbar4m2}
\end{center}
\end{figure}

\begin{center}
\begin{tabular}{l}
$A_1 \uparrow \hm{P}{\bar 4m2} \sim {{M_5}} \oplus \cdots$\\
$A_2 \uparrow \hm{P}{\bar 4m2} \sim {M_5}\oplus \cdots$\\
$B_2 \uparrow \hm{P}{\bar 4m2} \sim {M_1}\oplus{M_2}\oplus \cdots$\\
$B_1 \uparrow \hm{P}{\bar 4m2} \sim {M_3}\oplus{M_4}\oplus \cdots$\\
$\overline{E} \uparrow \hm{P}{\bar 4m2} \sim {\overline{M}_6}\oplus{\overline{M}_7}\oplus \cdots$\\
\end{tabular}
\end{center}
The obtained results coincide exactly with the corresponding data of Table 2 of ref. \onlinecite{kitaev1997}.
\end{enumerate}
\section{\label{sec:bandrepresentations}Band representations and Elementary band representations} 
\subsection{\label{bandrepproblem}The problem}
The concept of a band representation (BR) was introduced by \onlinecite{zak1982}, as a set of extended energy states over the entire reciprocal space, $E_n(\mathbf{k})$, related to the symmetry of (exponentially) localized states (Wannier orbitals).
The basic structure description of a crystal includes the assignation of a space group and the determination of the lattice parameters and of the atomic coordinates (occupied Wyckoff positions) of a minimal set of atoms that belong to the asymmetric unit (known also as the set of independent atoms). The atomic positions of all the atoms in the crystal are obtained through the action of the symmetry operations of the space group onto the coordinates of the independent atoms. As explained in Section \ref{sec:sitesymmetry}, the finite set of symmetry operations of the space group that keeps a point of a Wyckoff position invariant is its site-symmetry group $\gro{S}$, and it is isomorphic to a point group. It is important to note that the site-symmetry groups $\gro{S}_q$ of all points $q$ belonging to a Wyckoff position $\mathbf{Q}=\{q\}$ are isomorphic to the same point-group type, and in that sense one speaks of a site-symmetry group of a Wyckoff position $\gro{S}_{\mathbf{Q}}$. The localized states  (the atomic orbitals, for example) of an atom that occupy a given Wyckoff position transform according to a representation $\mathbf{d}_{\alpha}$ of its site-symmetry group. When the spin-orbit coupling is also considered, the localized states transform according to a representation of the double site-symmetry group, isomorphic to a double point group. A \emph{band representation} can be defined as the induced representation $\mathbf{d}_{\alpha}\uparrow\mathcal{G}$, being a particular case of the general description of the site-symmetry approach discussed in Section \ref{sec:sitesymmetry} If the representation $\mathbf{d}_{\alpha}$ of the site-symmetry group $\gro{S}$ is reducible, then it decomposes into irreps of the site-symmetry group,
\begin{equation}\label{ind-decband}
\mathbf{d}_\alpha= \bigoplus_\beta n_\beta^{(\alpha)} \mathbf{d}_\beta,
\end{equation}
and the induced BR decomposes into BRs induced from the irreducible representations $\mathbf{d}_\beta$
\begin{equation}\label{ind-decbandirrep}
\mathbf{d}_\alpha\uparrow\mathcal{G}= \bigoplus_\beta  n_\beta^{(\alpha)}(\mathbf{d}_\beta\uparrow\mathcal{G}).
\end{equation}
Therefore, in order to have a complete information of the BRs induced from a given Wyckoff position, it is necessary to calculate only the BRs induced from the \emph{irreps} of its site-symmetry group. 

The BRs induced from different irreps of the same site-symmetry group are not equivalent, but BRs induced from different Wyckoff positions could be equivalent. The definition of equivalence of BRs is different from the definition of equivalence of representations, where we say that two representations are equivalent if all the multiplicities of their decomposition into irreps are the same. However, for certain applications, as for example, the study of topological phases, we need a stronger form of equivalence \cite{michel1992,zeiner2000,NaturePaper,EBRtheory}. 

\textbf{Definition 1:} Two band representations $\rho_{\gro{G}}^{\mathbf{k}}$ and $\sigma_{\gro{G}}^{\mathbf{k}}$ are equivalent iff there exists a unitary matrix-valued function $S(\mathbf{k},s,\grel{g})$ smooth in $\mathbf{k}$ and continuous in $s$ such that for all $\grel{g}\in\mathcal{G}$
\begin{itemize}
\item[1.] $S(\mathbf{k},s,\grel{g})$ is a band representation for all $s\in[0,1]$,
\item[2.] $S(\mathbf{k},0,\grel{g})=\rho_{\gro{G}}^{\mathbf{k}}(\grel{g})$ and
\item[3.] $S(\mathbf{k},1,\grel{g})=\sigma_{\gro{G}}^{\mathbf{k}}(\grel{g})$
\end{itemize}
This definition implies that the BRs $\rho_{\gro{G}}^{\mathbf{k}}(\grel{g})$ and $\sigma_{\gro{G}}^{\mathbf{k}}(\grel{g})$ subduce into the same little group representations at all points in the Brillouin zone. This definition also implies the following: consider two Wyckoff positions $\mathbf{Q}$ and $\mathbf{Q}\,'$ and let $\rho_{\gro{G}}^{\mathbf{k}}(\grel{g})$ and $\sigma_{\gro{G}}^{\mathbf{k}}(\grel{g})$ be two BRs induced from an irrep $\rho$ of the site-symmetry group $\mathcal{S}_{\mathbf{Q}}$ of $\mathbf{Q}$ and an irrep $\sigma$ of the site-symmetry group $\mathcal{S}_{\mathbf{Q}\,'}$ of $\mathbf{Q}\,'$. The intersection $\mathcal{S}_{\mathbf{Q}_0}=\mathcal{S}_{\mathbf{Q}}\cap\mathcal{S}_{\mathbf{Q}\,'}$ of the two site-symmetry groups is the site-symmetry group of another Wyckoff position $\mathbf{Q}_0$ that can be identified relatively easy: some of the point coordinates of $\mathbf{Q}_0$ should be represented by variable parameters that interpolate between the point coordinates of the Wyckoff positions $\mathbf{Q}$ and $\mathbf{Q}\,'$. If for a given irrep $\rho_0$ of $\mathcal{S}_{\mathbf{Q}_0}$, the induced representations into the site-symmetry groups of $\mathbf{Q}$ and $\mathbf{Q}\,'$ satisfy, $\rho_0\uparrow S_{\mathbf{Q}}=\rho$ and $\rho_0\uparrow S_{\mathbf{Q}\,'}=\sigma$,then the two BRs $(\rho\uparrow G)\downarrow G^{\mathbf{k}}=\rho_G^{\mathbf{k}}$ and $(\sigma\uparrow G)\downarrow G^{\mathbf{k}}=\sigma_G^{\mathbf{k}}$ are equivalent.


Once a criterion of equivalence of BR is established, we define:

\textbf{Definition 2:} A band representation is called \emph{composite} if it is equivalent to the direct sum of other band representations. A band representation that is not composite is called \emph{elementary}.

All the elementary band representations (EBR) are induced from the so-called \emph{Wyckoff positions of maximal symmetry} (\emph{cf.} Section \ref{wyckoffmaximal} for the definition of the concept) but the opposite is not true. This fact was already pointed out for single-valued BRs in \onlinecite{bacry1988}, where also the complete list of exceptions was given. The list of exceptions for BRs induced from double-valued irreducible representations of the double space groups can be found in the supplementary material of \onlinecite{NaturePaper}.

The generalization of the procedure for the calculation of EBRs for systems with time-reversal symmetry is straightforward. In this case, it is enough to consider the BRs  induced from the physically irreducible representations of the site-symmetry groups of the Wyckoff positions of maximal symmetry. The physically irreducible or TR-invariant representations of the site-symmetry groups can be constructed following the steps explained in Section \ref{subsec:complexconjugation}. The TR-invariant irreps induce TR-invariant BRs in $\mathcal{G}$, and using the definitions 1 and 2 above the TR-invariant EBRs can be calculated. Not all the single-valued TR-invariant BRs induced from the double site-symmetry groups of Wyckoff positions are elementary: the list of exceptions is given in \cite{NaturePaper}. However, all the double-valued TR-invariant BRs induced from the double site-symmetry groups of Wyckoff positions of maximal symmetry are elementary, as a consequence of Kramer's theorem \cite{NaturePaper}.

In a series of concomitant articles \cite{EBRtheory,NaturePaper,GraphDataPaper}, in relation to the problem of topological insulators, we have studied the \emph{connectivities} of energy bands of the EBRs. The energy bands are continuous functions defined in the Brillouin zone. The possible different ways in which the Bloch states at bands of $\mathbf{k}$-\emph{vectors of maximal symmetry} (\emph{cf.} Section \ref{kmax}) are connected through intermediate bands of $\mathbf{k}$-vectors of lower symmetry, are restricted by the corresponding compatibility relations (\emph{cf.} Section \ref{sec:compatibilityrelations}). In most cases the EBRs are fully connected, which means that we can continuously 'travel' along all the points that form the energy band. However, in some cases, depending on the specific values of the electronic band energies at each $\mathbf{k}$-vector, the compatibility relations allow the decomposition of the whole group of bands into disconnected branches, separated by an energy gap. Exactly such EBRs describe topological insulators. In a companion paper \cite{GraphDataPaper} we describe in detail the algorithms used to determine if an EBR (with and without TR) is decomposable or not.

\subsection{\label{sec:methodbandrep}The method}
The algorithms implemented in the program {\tt BANDREP} to calculate the BRs follows the site-symmetry procedure explained in Section \ref{sec:sitesymmetry} and makes use of different tools of the BCS. Given a space group \gro{G}, a BR is fully identified by a Wyckoff position, an irrep of its site-symmetry group and the irreps of the little groups of any $\mathbf{k}$-vector of the reciprocal space. In fact, it is only necessary to consider the $\mathbf{k}$-vectors of maximal symmetry (\emph{cf.} Section \ref{kmax}), because the multiplicities of the irreps of the little groups of any other $\mathbf{k}$-vector can be calculated from the former ones using the corresponding compatibility relations described in Section \ref{sec:compatibilityrelations}. The multiplicities of the irreps of the little groups of any $\mathbf{k}$-vector in BRs are determined applying the site-symmetry approach.

Given a space group $\mathcal{G}$ and a Wyckoff position $\mathbf{Q}=\{q\}$, first the program identifies the symmetry elements included in the site-symmetry group $\gro{S}_q$: for each of the coset representatives $\{R|\mathbf{v}\}$ of $\gro{G}:\gro{T}$ the program calculates $\mathbf{t}$ that satisfies the equation,
\begin{equation}\label{sitesymchek}
\mathbf{t}=Rq+\mathbf{v}-q.
\end{equation}
If $\mathbf{t}$ belongs to the set of lattice translations of $\mathcal{G}$, then $\{R|\mathbf{v}-\mathbf{t}\}$ belongs to the site-symmetry group $\mathcal{S}_{q}$. Once all elements of $\mathcal{S}_{q}$ are determined, the tool {\tt IDENTIFY GROUP} (www.cryst.ehu.es/cryst/identify\_group) of the BCS is used to identify the point (or double point) group type isomorphic to $\mathcal{S}_{q}$ and to establish the corresponding isomorphic mapping between the group elements. As a next step, for each element $\grel{g}\in\mathcal{S}_q$, we calculate: (i) the characters $\chi_\alpha(\grel{g})$ of each irrep $\mathbf{d}_\alpha$ of the double point group (using {\tt Representations DPG}), and (ii) the characters $\chi_\beta^{^{*}\mathbf{k}}(\grel{g})$ of the irreps $D_\beta^{*\mathbf{k}}$ of $\mathcal{G}$ for each of the tabulated $\mathbf{k}$-vectors in the Brillouin zone, induced from the allowed irreps (or TR-invariant irreps) $\mathbf{d}_\beta^{\mathbf{k}}$ of the little group $\mathcal{G}^{\mathbf{k}}$ of $\mathbf{k}$ (applying the tool {\tt Representations DSG}).
 
The decomposition of the subduced representations $\mathbf{D}_\beta^{*\mathbf{k}}\downarrow\mathcal{S}_{q}$ into irreps $\mathbf{d}_\alpha$ of the site-symmetry group $\gro{S}_q$ is performed using eqs. (\ref{eq1}) and (\ref{eq2}):

\begin{equation}\label{eq1}
\mathbf{D}_\beta^{*\mathbf{k}}\downarrow\mathcal{S}_{q}\sim\bigoplus n_\alpha^{\mathbf{k},(\beta)}\mathbf{d}_\alpha,
\end{equation}
where the multiplicities are,
\begin{equation}\label{eq2}
n_\alpha^{\mathbf{k},(\beta)}=\frac{1}{\left|\mathcal{S}_q\right|}\sum_{\grel{g}\in\mathcal{S}_q}\chi_\alpha(\grel{g})^*\chi_\beta^{*\mathbf{k}}(\grel{g}).
\end{equation}

As explained in Section \ref{sec:sitesymmetry}, according to the Frobenius reciprocity theorem, these multiplicities coincide with the multiplicities of the irreps of the little group $\mathcal{G}^{\mathbf{k}}$ in the BR induced from the irrep $\mathbf{d}_\alpha$ of the double point group $\mathcal{S}_{q}$:

\begin{equation}\label{eq:multiplicities}
(\mathbf{d}_\alpha\uparrow \mathcal{G})\downarrow\mathcal{G}^{\mathbf{k}}\sim\bigoplus n_{\beta}^{\mathbf{k},(\alpha)}   \mathbf{d}_\beta^{\mathbf{k}}.
\end{equation}

We thus fully identify the BR induced from the irrep $\mathbf{d}_\alpha$ of the site-symmetry group of the Wyckoff position $\mathbf{Q}$ giving the tabulated irreps $\mathbf{d}_\beta^{\mathbf{k}}$ of the little group for every $\mathbf{k}$ of the space group $\mathcal{G}$ (obtained by {\tt REPRESENTATIONS DSG}) and calculate the multiplicities $n_{\alpha}^{\mathbf{k},(\beta)}$.

Finally, we check if the BR induced from the Wyckoff position $\mathbf{Q}$ is elementary or not. As all EBRs of a space group are induced from Wyckoff positions of maximal symmetry, the check is reduced to BRs induced from such Wyckoff positions. Each BR is characterized by a set of multiplicities of the little-group irreps but only for $\mathbf{k}$-vectors of maximal symmetry. Using these lists of multiplicities we can easily check if the given BR decomposes at every \textbf{k} as the direct sum of two or more irreps. If such decompositions exist, the BR is a candidate to be a composite BR. 

However, as explained in Section \ref{sec:bandrepresentations}, this condition is not sufficient to consider the BR as equivalent to the direct sum of BRs induced from a different Wyckoff position $\mathbf{Q}\,'$. In addition, it is necessary to calculate the intersection of the site-symmetry groups $\gro{S}_{\bf{Q}}$ and $\gro{S}_{\bf{Q}}\,'$ of $\mathbf{Q}$ and $\mathbf{Q}\,'$, respectively, and identify the Wyckoff position $\mathbf{Q}_0$ (of non-maximal symmetry) which has this intersection $\gro{S}_{\bf{Q}_0}$ as its site-symmetry group. Then, we induce representations of $\gro{S}_{\bf{Q}}$ and $\gro{S}_{\bf{Q}}\,'$ from every irrep of $\gro{S}_{\bf{Q}_0}$. If for some irrep of $\gro{S}_{\bf{Q}_0}$ we get $\rho$, the irrep of $\gro{S}_{\bf{Q}}$ that induces the candidate BR to be composite, and a reducible representation $\sigma$ of $\gro{S}_{\bf{Q}}\,'$, that induces a composite BR that decomposes at every $\mathbf{k}$ as $\rho\uparrow\gro{G}$ does, the candidate BR is composite. Otherwise, it is elementary.


\subsection{\label{wyckoffmaximal}Wyckoff positions of maximal symmetry}
The so-called \emph{Wyckoff positions of maximal symmetry} are of importance for certain applications, and in particular, for the calculation of the EBRs. In the following, we comment briefly on the definition of Wyckoff positions of maximal symmetry The intersection $\mathcal{S}_{\mathbf{Q}_0}=\mathcal{S}_{\mathbf{Q}}\cap\mathcal{S}_{\mathbf{Q}\,'}$ of the two site-symmetry groups is the site-symmetry group of another Wyckoff position $\mathbf{Q}_0$ that can be identified relatively easy: some of the point coordinates of $\mathbf{Q}_0$ should be represented by variable parameters that interpolate between the point coordinates of the Wyckoff positions $\mathbf{Q}$ and $\mathbf{Q}\,'$. If for a given irrep $\rho_0$ of $\mathcal{S}_{\mathbf{Q}_0}$, the induced representations into the site-symmetry groups of $\mathbf{Q}$ and $\mathbf{Q}\,'$ satisfy, $\rho_0\uparrow S_{\mathbf{Q}}=\rho$ and $\rho_0\uparrow S_{\mathbf{Q}\,'}=\sigma$, then the two BRs $(\rho\uparrow G)\downarrow G^{\mathbf{k}}=\rho_G^{\mathbf{k}}$ and $(\sigma\uparrow G)\downarrow G^{\mathbf{k}}=\sigma_G^{\mathbf{k}}$ are equivalent.and on the procedure we used for their calculation. Note that the sets of Wyckoff positions of double space groups and those of the corresponding space groups are closely related. In fact, the essential difference concerns the site-symmetry groups which, in the former case, are isomorphic to double point groups, while in the latter, only to point groups. As a consequence, practically the same procedure for the determination of the Wyckoff positions of maximal symmetry can be used for space groups and for double space groups.

It is common to describe a Wyckoff position by its multiplicity, Wyckoff letter, symbol of the site-symmetry group and a set of coordinate triplets of the points in the unit cell that belong to the Wyckoff position, possibly depending on one or two variable parameters (three for the general position) (for a detailed introduction to Wyckoff positions of space groups, \emph{cf.} \ita). A Wyckoff position $\bf{Q}$ with a site-symmetry group $\gro{S}_{\bf{Q}}$ has maximal symmetry if it is not \emph{connected} to another Wyckoff position $\bf{Q}\,'$ whose site-symmetry group $\gro{S}_{\bf{Q}\,'}$ is a supergroup of $\gro{S}_{\bf{Q}}$. We say that two Wyckoff positions are \emph{connected} if (i) the coordinate triplet of at least one of them depends on one or more variable parameters, and (ii) if for specific values of the variable  parameters the coordinate triplets of the two Wyckoff positions coincide. For instance, in the space group P$2/m$ (No. 10), the site-symmetry group of the Wyckoff position $2i: (0,y,0)$ is isomorphic to the point group $2$. It is not a Wyckoff position of maximal symmetry because it is \emph{connected} to the Wyckoff position $2a: (0,0,0)$:  the coordinate-triplet description of $2a$ is obtained from that of $2i$ by the substitution $y=0$, and the site-symmetry group of the position $2a$ is isomorphic to $2/m$ which is supergroup of $2$.

The algorithm to identify the Wyckoff positions of maximal symmetry is straightforward. In the space groups with no points in a special position (the so-called \emph{fixed-point-free} or \emph{Bieberbach} groups) the general Wyckoff position is the only Wyckoff position of maximal symmetry. For the rest of space groups, we distribute the special Wyckoff positions into three subsets, according to the number $(0,1,2)$ of variable parameters of their coordinate triplets. All Wyckoff positions with $0$ variable parameters are Wyckoff positions of maximal symmetry. Those Wyckoff positions of the subset with one variable parameter which are \emph{connected} to at least one Wyckoff position of the subset with no variable parameter, in the sense explained above, are not maximal. The rest of Wyckoff positions with one variable parameter are maximal. Finally, we repeat the check for the subset of Wyckoff positions with two variable parameters, trying to find if they are \emph{connected} to at least one of the Wyckoff positions of the subsets with 0 and 1 variable parameter. Those that have no \emph{connection} are Wyckoff positions of maximal symmetry.

As an example, consider the Wyckoff positions of the (double) space group P$4/ncc$ (No. 130) shown in Table \ref{table:wyckoffmax}.  The Wyckoff positions $4a,\, 4b,\,4c$ and $8d$  are of maximal symmetry while $8e,\, 8f$ and $16g$ are not. If a Wyckoff position is not maximal, then the \emph{corresponding} Wyckoff position of maximal symmetry (\emph{i.e.} the one to which it is \emph{connected}) together with the specific values of the variable parameters for which the two coordinate-triplet descriptions coincide, are indicated in the last column of the table. 

\begin{table}
\caption{Table 1: Wyckoff positions of the (double) space group P$4/ncc$ (No. 130). Each Wyckoff position is specified by its multiplicity and Wyckoff letter (first column), the Hermann-Mauguin (Sch\"onflies) symbol of its site-symmetry group (second column), the coordinate-triplet description of a representative of the orbits of the Wyckoff positions (third column). In the last column it is indicated if the Wyckoff position is of maximal symmetry or not. For the Wyckoff positions of non-maximal symmetry, the \emph{corresponding} Wyckoff position of maximal symmetry, together with the specific values of the variable parameters for which the two coordinate-triplet descriptions coincide, are also specified.}
\label{table:wyckoffmax}

\begin{tabular}{cccc}
Wyckoff&Site-symmetry&coordinate&maximal\\
position&group&triplet&symmetry\\
\hline
$4a$&$222(D_2)$&$(3/4,1/4,1/4)$&yes\\
$4b$&$\bar{4}(S_4)$&$(3/4,1/4,0)$&yes\\
$4c$&$4(C_4)$&$(1/4,1/4,z)$&yes\\
$8d$&$\bar{1}(C_i)$&$(0,0,0)$&yes\\
$8e$&$2(C_2)$&$(3/4,1/4,z)$&no\\
&&&$z=1/4\to 4a$\\
$8f$&$2(C_2)$&$(x+1/2,x,1/4)$&no\\
&&&$x=1/4\to 4a$\\
$16g$&$1(C_1)$&$(x,y,z)$&no\\
&&&$x=y=z=0\to 8d$
\end{tabular}
\end{table}

\subsection{\label{kmax}$\mathbf{k}$-vectors of maximal symmetry}
In analogy to the Wyckoff positions of maximal symmetry introduced in the previous section, for each space group we can define a set of $\mathbf{k}$-\emph{vectors of maximal symmetry}. These vectors play also an important role, for example, in the analysis of the connectivities of the BRs (\emph{cf.} \cite{GraphTheoryPaper,GraphDataPaper,EBRtheory}). 

Similar to the distribution of points of direct space into Wyckoff positions, the set of all $\mathbf{k}$-vectors can be distributed into the so-called $\mathbf{k}$-\emph{vector types}. A $\mathbf{k}$-vector type consists of complete orbits of $\mathbf{k}$-vectors and thus of full stars of $\mathbf{k}$-vectors. The $\mathbf{k}$-vectors belonging to a $\mathbf{k}$-vector type are represented by a $\mathbf{k}$-vector letter (here we follow CDML notation), by the point-group type of the little co-groups of $\mathbf{k}$-vectors, and by a set of $\mathbf{k}$-vectors coefficients: the zero, one or two variable parameters in the $\mathbf{k}$-vector coefficients correspond to special  $\mathbf{k}$-vector types, \emph{i.e.}, they define symmetry points, symmetry lines or symmetry planes in the Brillouin zone. Three variable parameters indicate a general  $\mathbf{k}$-vector type. 
We say that a $\mathbf{k}$-vector type (or just a $\mathbf{k}$-vector, for short) is \emph{of maximal symmetry} if its little co-group is not a subgroup of the little co-group of another $\mathbf{k}\,'$-vector type \emph{connected} to $\mathbf{k}$ in the same sense as the \emph{connected} Wyckoff positions discussed in Section \ref{wyckoffmaximal}. The procedure to identify the $\mathbf{k}$-vectors of maximal symmetry is analogous to the procedure for the determination of the Wyckoff positions of maximal symmetry. 

In general, the set of $\mathbf{k}$-vectors of maximal symmetry for non-centrosymmetric space groups (\emph{i.e} space groups \gro{G} whose point groups $\overline{\gro{G}}$ do not include the operation of (space) inversion $\overline{1}$) could be modified when time-reversal symmetry is taken into account. One can show that in such cases wave vectors of the same $\mathbf{k}$-vector type (with respect to the spacial symmetry) could behave differently under the action of time reversal: some wave vectors are time-reversal invariant, the so-called \emph{Time Reversal Invariant Momentum} (TRIM) points, while others are not. In other words, when TR symmetry is taken into account, the TRIM points are $\mathbf{k}$-vectors of maximal symmetry and the corresponding physically irreducible representations  determine the transformation properties of the eigenfunctions of the Hamiltonian of the system. As an example, consider the polar space group $\hm{P}{4}$ (No. 75) and the symmetry line $\Lambda\,(0,0,w)$ which is a $\mathbf{k}$-vector line of maximal symmetry as its little co-group is the point group 4 ($C_4$). There are 4 one-dimensional single-valued irreps at the $\Lambda\,(0,0,w)$ point: $\Lambda_i$, $i=1,\ldots,4$. If however, time-reversal symmetry is taken into account, the $\mathbf{k}$-vector type $\Lambda\,(0,0,w)$ 'splits' into three $\mathbf{k}$-vector types: the points $\Gamma\, (0,0,0)$ and $Z\,(0,0,1/2)$ are TRIM points and become $\mathbf{k}$-vector types of maximal symmetry while the rest of the points of the line $\Lambda$ form a $\mathbf{k}$-vector type of \emph{non}-maximal symmetry. Two of the four single-valued irreducible representations at these points, $\Gamma_3$, $\Gamma_4$ and $Z_3$, $Z_4$, form a pair of complex conjugated irreps and become doubly-degenerate. 
Tables \ref{table:kvecsnoTR} and \ref{table:kvecsTR} show the lists of $\mathbf{k}$-vectors of maximal symmetry of $\hm{P}{4}$ (No. 75), without and with TR, respectively.

The above example indicates the important consequences of the time-reversal symmetry in the derivation of the compatibility relations, and in particular, in the study of EBR connectivities between pairs of $\mathbf{k}$-vectors of maximal symmetry.
  
\begin{table}
\caption{Table 2: $\mathbf{k}$-vectors of maximal symmetry of the (double) space group P$4$ (No. 75) if time-reversal symmetry is not considered. The $\mathbf{k}$-vector labels follow the notation of CDML. The second column gives the little co-group of the $\mathbf{k}$-vector. The third column shows the coefficients of a representative of the $\vec{k}$-vector star.}
\label{table:kvecsnoTR}

\begin{tabular}{ccc}
label&little co-group&coefficients\\
\hline
$\Lambda$&$4(C_4)$&$(0,0,w)$\\
$W$&$4(C_4)$&$(0,1/2,w)$\\
$V$&$4(C_4)$&$(1/2,1/2,w)$
\end{tabular}
\end{table}

\begin{table}
\caption{Table 3: $\mathbf{k}$-vectors of maximal symmetry of the (double) space group P$4$ (No. 75) with time-reversal symmetry taken into account. The $\mathbf{k}$-vector labels in the first column follow the notation of CDML. The second column gives the little co-group of the $\mathbf{k}$-vector. The third column shows the coefficients of a representative of the $\vec{k}$-vector star.}
\label{table:kvecsTR}

\begin{tabular}{ccc}
label&little co-group&coefficients\\
\hline
$\Gamma$&$4(C_4)$&$(0,0,0)$\\
$Z$&$4(C_4)$&$(0,0,1/2)$\\
$X$&$4(C_4)$&$(0,1/2,0)$\\
$R$&$4(C_4)$&$(0,1/2,1/2)$\\
$M$&$4(C_4)$&$(1/2,1/2,0)$\\
$A$&$4(C_4)$&$(1/2,1/2,1/2)$
\end{tabular}
\end{table}

\subsection{\label{bandrep}The program {\tt BANDREP}}
The program {\tt BANDREP}, recently added to BCS, applies the method explained in Section \ref{sec:methodbandrep} to calculate the BRs of any of the 230 double space groups. The fact that the list of single-valued BRs identified as elementary by the program, coincides exactly with the list obtained by \onlinecite{bacry1988}, alongside with the many checks performed and the agreement between the program and the results of \onlinecite{NaturePaper}, can be considered a proof test of the program.

\begin{figure}
\begin{center}
\includegraphics[width=\textwidth]{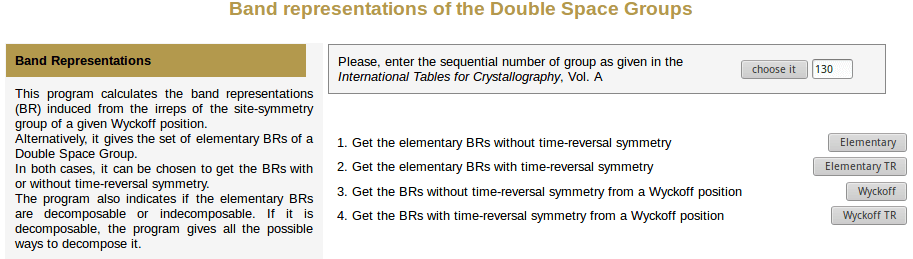}
\caption{Figure 14: Screenshot of the first input form of the program {\tt BANDREP}. The input data include (i) specification of the space group, and (ii) choice among the four options: (1) elementary band representations (EBR) without time-reversal (TR); (2) EBRs with TR; (3) band representations (BR)  without TR; and (4) BRs with TR.}
\label{fig:bandrepinput1}
\end{center}
\end{figure}

\begin{figure}
\begin{center}
\includegraphics[width=0.7\textwidth]{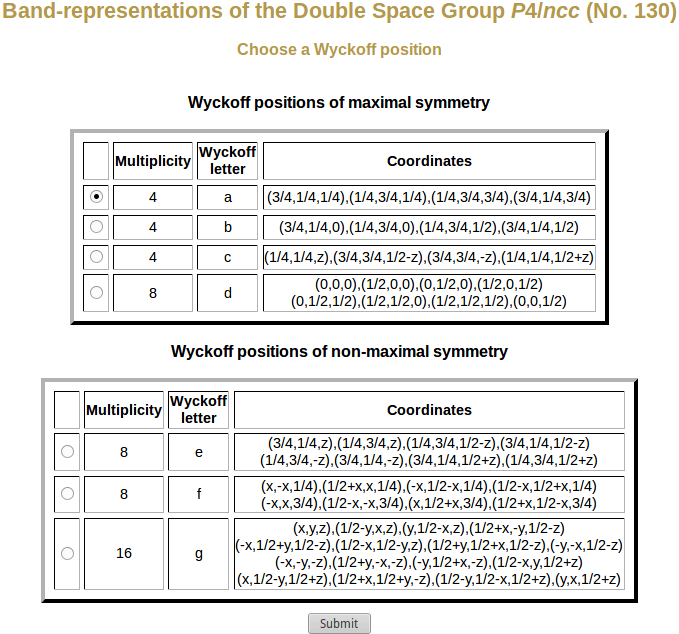}
\caption{Figure 15: Screenshot of the second input form of the program {\tt BANDREP} which shows the set of Wyckoff positions of the specified space group divided into positions of maximal and non-maximal symmetry. This input form is shown only for the options (3) and (4) of the main input (see figure \ref{fig:bandrepinput1}). }
\label{fig:bandrepinput2}
\end{center}
\end{figure}

\paragraph{Input:}
The main page of the input requires the specification of double space group (by its \ita sequential number) and offers four options: calculation either of EBRs of the double space group with or without TR symmetry, or of the BRs with or without TR symmetry. For the last two options the program produces the list of Wyckoff positions of the selected space group, separated into those of maximal and non-maximal symmetry. Figures \ref{fig:bandrepinput1} and \ref{fig:bandrepinput2} show screenshots of the input pages.

\paragraph{Output:}
The option selected by the user determines the specific output produced by the program:
\begin{enumerate}
\item Option: \emph{Elementary band representations without TR symmetry} 

A screenshot of the table output of the program {\tt BANDREP} for the double space group $\hm{P}{4/ncc}$ (No. 130) is shown on Figure \ref{fig:bandrepoutput1}. The EBRs are listed in columns specified by (i) the Wyckoff positions of $\hm{P}{4/ncc}$ (No. 130) followed by the symbol of the double point group $\gro{S}_{\mathbf{Q}}$ isomorphic to the site-symmetry group (first row of the header), and (ii) the irrep $\mathbf{d}_\alpha$ of the site-symmetry group from which the  EBR $\mathbf{d}_\alpha\uparrow \gro{G}$ is induced (second row of the header); the dimension of $\mathbf{d}_\alpha\uparrow \gro{G}$ is shown in brackets after the EBR symbol. The entries of the third row of the header indicate if the EBR is decomposable or not (\emph{cf.} Section \ref{bandrepproblem}). The entries of the output table show the decompositions of EBRs into irreps of the little groups of the $\mathbf{k}$-vectors of maximal symmetry which denote the rows of the table. The dimensions of the little-group irreps are given in brackets after their symbols.
\begin{figure}
\begin{center}
\includegraphics[width=\textwidth]{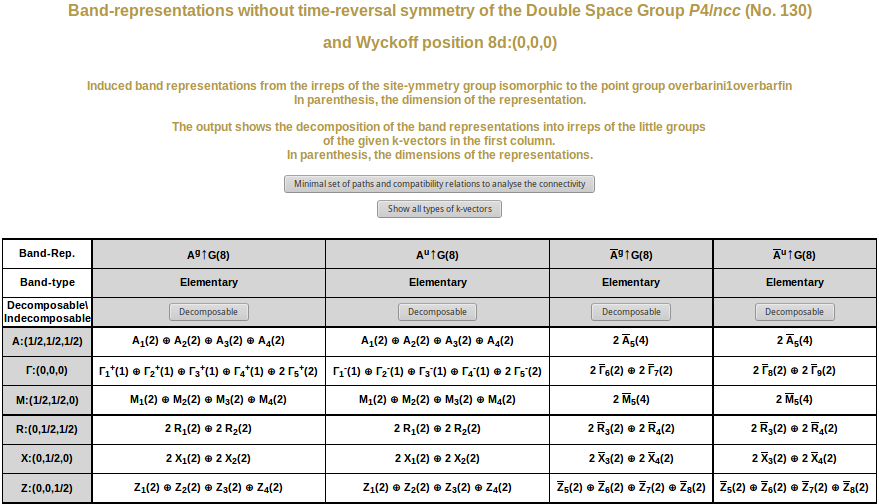}
\caption{Figure 16: Screenshot of the output given by the program {\tt BANDREP} for $P4/ncc$ (No. 130) and for the option 'elementary band representations without time-reversal symmetry' (\emph{cf.} Fig. \ref{fig:bandrepinput1}). For a detailed description of the displayed data, see Section \ref{bandrep}. Only part of the wider output has been included in the figure.}
\label{fig:bandrepoutput1}
\end{center}
\end{figure}

Clicking on the button "Show all types of $\mathbf{k}$-vectors" generates a table with the decompositions of each EBR into irreps of the little groups of maximal and non-maximal $\mathbf{k}$-vectors.

Clicking on the button "Minimal set of paths and compatibility relations to analyse the connectivity", produces the set of paths between pairs of $\mathbf{k}$-vectors of maximal symmetry to be considered in the analysis of the connectivities of the EBRs in the given space group, once the known redundancies have been removed \cite{GraphDataPaper}, and the corresponding independent compatibility relations. Figure \ref{fig:paths1} shows the screenshot of the minimal set of paths and compatibility relations for the space group $\hm{P}{4/ncc}$. For more details about the problem of connectivity of EBRs see \onlinecite{GraphDataPaper}.
\begin{figure}
\begin{center}
\includegraphics[width=\textwidth]{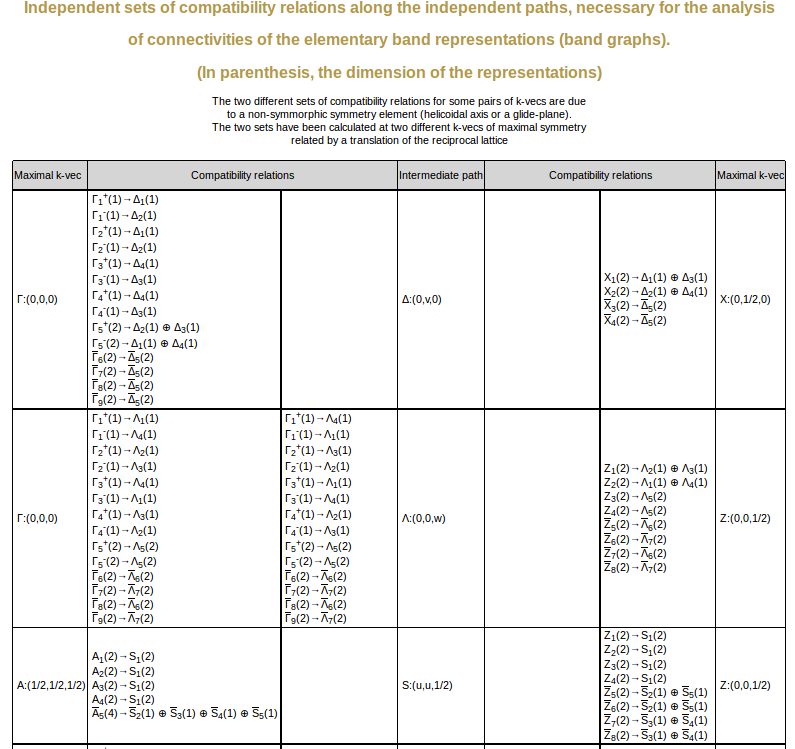}
\caption{Figure 17: Screenshot of the output given by the program {\tt BANDREP} which shows the independent sets of compatibility relations along intermediate paths between the $\mathbf{k}$-vectors of maximal symmetry of $P4/ncc$ (No. 130), necessary for the analysis of the connectivity structure of elementary band representations. Only part of the longer output has been included in the figure.}
\label{fig:paths1}
\end{center}
\end{figure}

\item Option: \emph{Elementary band representations with TR symmetry} 
 
The essential difference in comparison with the output of Option 1 is that the shown results refer to \emph{physically-irreducible} representations (for a discussion on physically-irreducible representations, \emph{cf.} Section \ref{subsec:complexconjugation}).
 
\item Option: \emph{Band representations from a Wyckoff position without TR symmetry}

Under this option the program shows the band representations for a specific Wyckoff position chosen from a list of Wyckoff positions of maximal or of non-maximal symmetry. Fig. \ref{fig:bandrepoutput2} shows the output produced by the program for the case of the double space group $\hm{P}{4/ncc}$, and the Wyckoff position of maximal symmetry $8d$. The structure of the output table is similar to those of elementary bands: the essential difference is that the program shows not only the elementary but also the composite band representations (specified in the second row of the table header). Whether the EBRs are decomposable or not is also indicated. In the partial output shown in Fig. \ref{fig:bandrepoutput2} all the EBRs are decomposable. By clicking on the button "Decomposable", we obtain all different ways to decompose the EBR into disconnected branches, which correspond to topological insulators. The four possible ways to decompose the EBR $A_g\uparrow \gro{G}$ induced from the $A_g$ irreducible representation of the site-symmetry group of the $8d$ Wyckoff position of the $\hm{P}{4/ncc}$  space group are shown in Fig. \ref{fig:bandrepoutputdesc}. Each column gives the arrangements of the set of the little-group irreps at each $\mathbf{k}$-vector of maximal symmetry in a branch. The band-graph (not yet available in the BCS) that illustrates a disconnected EBR is shown in Fig. \ref{fig:bandgraph}. The arrangements of irreps at each $\mathbf{k}$-vector in Fig. \ref{fig:bandgraph} correspond to the decomposition shown on the first row of Fig. \ref{fig:bandrepoutputdesc}.
\begin{figure}
\begin{center}
\includegraphics[width=\textwidth]{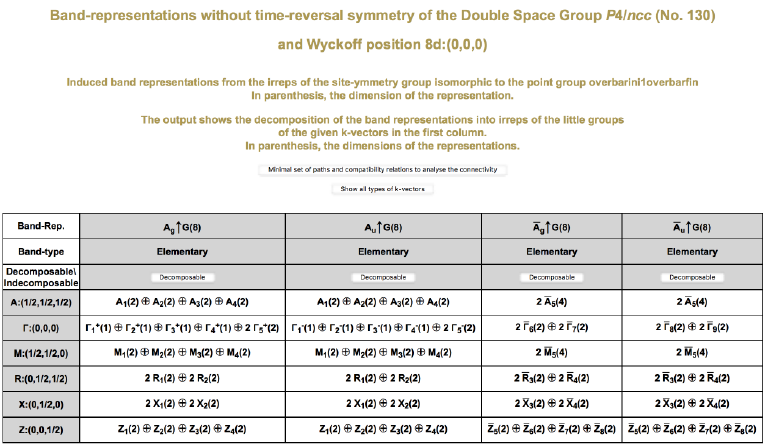}
\caption{Figure 18: Screenshot of the output given by the program {\tt BANDREP} which shows the band representations of $P4/ncc$ (No. 130) and Wyckoff position $8d$. For a detailed description of the displayed data, see Section \ref{bandrep}}
\label{fig:bandrepoutput2}
\end{center}
\end{figure}
\begin{figure}
\begin{center}
\includegraphics[width=0.6\textwidth]{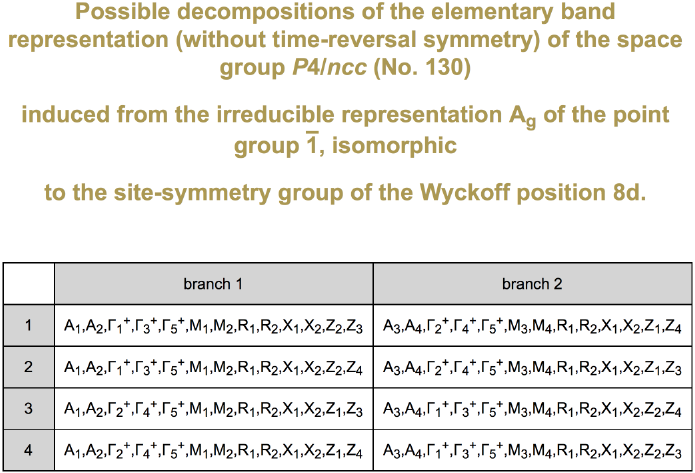}
\caption{Figure 19: Screenshot of the output given by the program {\tt BANDREP} which shows all (four) different ways to decompose the elementary band representation induced from the $A_g$ representation of the site-symmetry group of the  Wyckoff position $8d$ of $P4/ncc$ (No. 130). The elementary band representation can be decomposed into two further indecomposable branches (branch 1 and branch 2 in the figure). Each column of the table gives the arrangements of the set of the little-group irreps at each $\mathbf{k}$-vector of maximal symmetry in a branch.}
\label{fig:bandrepoutputdesc}
\end{center}
\end{figure}
\begin{figure}
\begin{center}
\includegraphics[width=\textwidth]{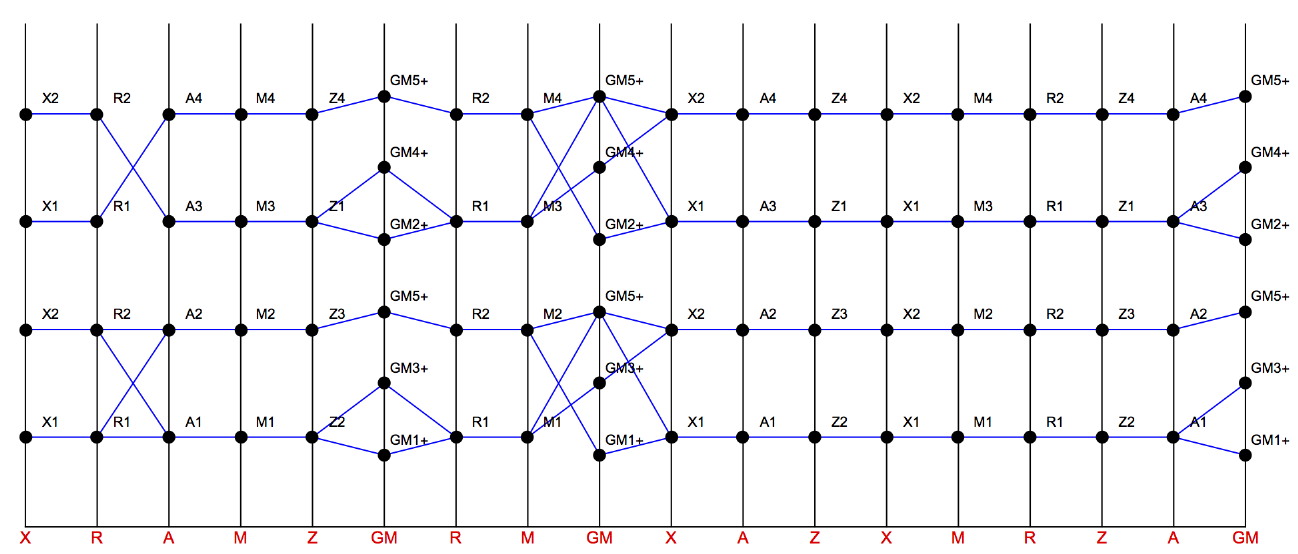}
\caption{Figure 20: Band-graph that illustrates a disconnected elementary band representation induced from the $A_g$ representation of the site-symmetry group of the Wyckoff position $8d$ of $P4/ncc$ (No. 130). The arrangements of irreps at each $\mathbf{k}$-vector correspond to the decomposition shown in the first row of Fig. \ref{fig:bandrepoutputdesc}. The interconnections between the little-group irreps of neighboring $\mathbf{k}$-vectors are drawn in accordance to the corresponding compatibility relations.}
\label{fig:bandgraph}
\end{center}
\end{figure}

\item Option: \emph{Band representations from a Wyckoff position with TR symmetry}. This provides similar information as Option 3 but for the \emph{physically-irreducible} representations.
\end{enumerate}

\section{Conclusions}

The group-theoretical description of physical systems where the Hamiltonian depends on spin components, would require the use of the so-called \emph{double crystallographic groups} and their single- and double-valued representations. In this paper, we describe a set of databases and programs of double crystallographic groups that recently have been implemented in the Bilbao Crystallographic Server (http://www.cryst.ehu.es). As the rest of the programs on BCS, the new tools are freely available and can be accessed via user-friendly web interfaces. Some of the algorithms applied in the new programs are extensions of the algorithms used in the BCS for ordinary space groups.

The tool {\tt DGENPOS} provides in different formats the symmetry operations of the 230 double space groups in the standard or conventional setting. The program {\tt REPRESENTATIONS DPG} gives access to the irreducible representations of the 32 crystallographic double point groups while {\tt REPRESENTATIONS DSG} calculates the irreducible representations of the double space groups and analyses their \emph{reality} indicating also the pairs of conjugated irreducible representations. 
The program {\tt DSITESYM} applies the site-symmetry approach to the double space groups. It can be considered as a bridge between a local description of the atomic orbitals on-site and a global description through extended states along the Brillouin zone in a structure. The program {\tt DCOMPREL} calculates the compatibility relations between the irreducible representations of double space groups at high- and low-symmetry points in the Brillouin zone, necessary in the analysis of the connectivity of the functions defined in the reciprocal space (\emph{e.g.} in the analysis of the structure of the electronic bands and their connectivity through the Brillouin zone). As an application of the site-symmetry approach, we have developed the program {\tt BANDREP} that provides the band representations and the subset of elementary band representations induced from any Wyckoff position of any double space group. The tool also identifies the subset of decomposable elementary band representations and the different ways of their decomposition, together with large amount of additional data necessary for their study. Concomitantly with the current paper, the results provided by {\tt BANDREP} have been successfully applied in a novel method for the description, search and prediction of topological insulators \cite{NaturePaper,GraphTheoryPaper,GraphDataPaper,EBRtheory}.

%

\appendix
\section{Normal-subgroup induction procedure}
The irreps of a space group ${\cal{G}}$ are obtained by induction from the irreps of its translation group ${\cal{T}}$. Assuming the Born-von Karman (periodic) boundary conditions
$(\textbf{I}, \textbf{t}_i)^{N_i} = (\textbf{I}, \textbf{o})$
to hold, where \textbf{t}$_i$ = (1,0,0), (0,1,0) or (0,0,1) and $N_i$ is a large integer for \textit{i} = 1, 2 or 3, respectively. Then, the irreps $\mathbf{\Gamma}^{\mathbf{k}}$ of the translation group ${\cal{T}}$ are given by:
\begin{equation}
\mathbf{\Gamma}^\textbf{k} [ (\textbf{I},\textbf{t})] = \mathrm{exp}(-i\textbf{k}\cdot\textbf{t}).
\end{equation} 
There are $N_1N_2N_3$ different irreps of $\gro{T}$ which are 
distinguished by the wave vectors:  
\begin{equation} \label{eqkve}
\mathbf{k} = \sum_{i=1}^3 k_{i}\,{\bf a}_{i}^{*}, 
\end{equation}
where $ k_{i} = q_i/N_i;\ q_i=0,\,1,\,2,\,\ldots\,,\,N_i-1$.
The basis \textbf{a}$_1^{*}$, \textbf{a}$_2^{*}$, \textbf{a}$_3^{*}$ 
is called the basis of the \emph{reciprocal lattice} \textbf{L}$^*$ and 
it is the \emph{dual basis} of $\mathbf{a}_1$,  
\textbf{a}$_2$,  \textbf{a}$_3$ of \textbf{L}. The vectors $\mathbf{a}_i^*$ 
are defined by the relations $\mathbf{a}_i\cdot\mathbf{a}_j^*=2\pi\delta_{ij}$ 
where $\delta_{ij}$ is the Kronecker symbol.

The wave vectors \textbf{k} and \textbf{k'} = \textbf{k} + \textbf{K}, where \textbf{K} is a vector of the reciprocal lattice \textbf{L}$^*$, describe the same irreps of $\cal{T}$. Therefore, in order to determine all the irreps of ${\cal{T}}$, it is necessary to consider only the \textbf{k}-vectors of the first Brillouin-zone.

The \textit{little co-group} of \textbf{k} is the point group consisting of all the rotational parts $R^\textbf{k}$ of the symmetry operations of the space group ${\cal{G}}$ that either leave the \textbf{k}-vector invariant, or map it to an equivalent vector, \emph{i.\,e.}:
\begin{equation}
{\bf k} = {\bf k}\bfit{W}^{\mathbf{k}} + {\bf K,~~ K} \in {\bf L}^*. \label{def_little_cogroup}
\end{equation}
 Here, $\bfit{W}^{\mathbf{k}}$ is the $(3\times3)$ matrix representation of $R^\textbf{k}$. The little co-group $\overline{{\cal{G}}}^\textbf{k}$ is a subgroup of the point group $\overline{{\cal{G}}}$ of the space group ${\cal{G}}$. The vector  {\bf k} is called a \emph{general} \textbf{k} \emph{vector} if the little co-group contains the identity operation only, \emph{i.e.} $\overline{\cal{G}}\,^{\mathbf{k}} = \{\cal{I}\}$; otherwise $\overline
{\cal{G}}\,^{\mathbf{k}} > \{\cal{I}\}$, and  {\bf k} is called a \emph{special} 
\textbf{k} \emph{vector}. 

\vspace{1ex} Consider the coset decomposition of $\overline{\gro{G}}$ relative to $\overline{\gro{G}}\,^{\mathbf{k}}$.
If $\{R_m\}$ is the corresponding set of coset representatives, then the set $\mathbf{*k}=\{{\bf k}\bfit{W}_m+\mathbf{K}\}$, with $\mathbf{K}\in$\textbf{L}$^*$, is called the \textit{star of {\bf k}} and the vectors $\mathbf{k}\bfit{W}_m+\mathbf{K}$ are called the \textit{arms of} $\mathbf{*k}$.

An orbit of $\mathbf{\Gamma}^{\mathbf{k}}(\gro{T})$ under conjugation by \gro{G} comprises 
all irreps $\mathbf{\Gamma}^{\mathbf{k}'}(\gro{T})$ with $\mathbf{k}'$ 
belonging to $\mathbf{*k}$. Irreps of \gro{T} belonging to the same orbit give rise to equivalent irreps of
\gro{G}, \emph{i.e.} in order to obtain each irrep of \gro{G} exactly once it is necessary to consider one \textbf{k} vector per star. (A simply connected subset of the Brillouin zone which contains 
exactly one \textbf{k} vector per  $\mathbf{*k}$, is called a 
\emph{representation domain}.)

Given a space group \gro{G}, its translation subgroup \gro{T}, and an irrep 
$\mathbf{\Gamma}^{\mathbf{k}}(\gro{T})$, one can define the \emph{little 
group} $\gro{G}^{\mathbf{k}}$ of the wave vector \textbf{k}: it is a space group 
that consists of all those elements of \gro{G} whose rotation parts 
$R^{\mathbf{k}}$ leave either \textbf{k} unchanged or 
transform it into an equivalent vector,

\begin{equation}
\gro{G}^{\mathbf{k}}=\{\{R^{\mathbf{k}}|\mathbf{v}^{\mathbf{k}}\} \in 
\gro{G}|R^{\mathbf{k}} \in \overline{\gro{G}}\,^{\mathbf{k}}\}.\label{litgr} 
\end{equation} 

The irreps of space groups are obtained by induction from the so-called \textit{allowed}
irreps of the little groups $\gro{G}^{\mathbf{k}}$ of \textbf{k}. If 
$\bfit{D}^{\mathbf{k},i}$ is an 
allowed irrep of $\gro{G}^{\mathbf{k}}$, then $\bfit{D}^{\mathbf{k},i}
(\{1|\,\mathbf{t}\})=\exp{(-i\,{\bf k\cdot\,t})}\,\bfit{I}$. 
(The matrix \bfit{I} is the identity matrix with
 $\dim\bfit{I}=\dim\bfit{D}^{\mathbf{k},\,i}$). 

The allowed irreps of the little group $\gro{G}^{\mathbf{k}}$ are determined by an induction procedure \cite{zak1960} which is based on the fact that the little groups (as all crystallographic groups) are \emph{solvable} groups, \emph{i.e.} for each group $\gro{G}^{\mathbf{k}}$ there exists a series of subgroups $\gro{H}_i^{\mathbf{k}}$ (the so-called \emph{composition series}), such that:

\begin{equation}\label{comp_little}
\gro{G}^{\mathbf{k}}\rhd\gro{H}_1^{\mathbf{k}}\rhd\ldots\rhd\gro{H}_{m-1}^{\mathbf{k}}
\rhd\gro{H}_{m}^{\mathbf{k}}\rhd\ldots\rhd\gro{H}_n^{\mathbf{k}}={\gro{T}}
\end{equation}
and that the factor groups $\gro{H}_{m-1}^{\mathbf{k}}/\gro{H}_{m}^{\mathbf{k}}$ are cyclic groups of prime order. 
The (allowed) irreps of $\gro{G}^{\mathbf{k}}$ can be obtained from the (allowed) irreps of 
${\gro{T}}$ by applying several times the general induction procedure 
'climbing up' the chain of normal subgroups (eq. \ref{comp_little}). Important for the irrep calculation is the observation that the factor groups in the composition series of crystallographic groups have orders 2 or 3 which simplifies considerably the induction procedure. The corresponding induction formulae and a detailed example of application of the induction method in the case of the space group $P4bm$ and $\mathbf{k}=\mathbf{X}(0,1/2,0)$ can be found, for example, in Aroyo \emph{et al.} (2006).

Finally, following the normal-subgroup induction procedure, the irreps of a space group \gro{G} (the so-called \emph{full} irreps) for a given \textbf{k} vector are obtained
 by induction from the allowed irreps $\mathbf{D}^{\mathbf{k},\,i}$ of 
the corresponding little group $\gro{G}^{\mathbf{k}}$. Let the elements 
$\grel{q}_m=\{R_m|\,\mathbf{v}_m\},\ \ m=1,\ \ldots\ ,s$ be the  
representatives of the cosets in the decomposition of \gro{G} relative to 
$\gro{G}^{\mathbf{k}}$:

\begin{equation}\label{coset}
{\cal{G}} = {\cal{G}}^{\mathbf{k}} \cup \grel{q}_2{\cal{G}}^{\mathbf{k}}\cup \dots \cup \grel{q}_s{\cal{G}}^{\mathbf{k}}.
\end{equation}

 If $dim\,\bfit{D}^{\mathbf{k},\,i}=r$, 
and if $s$ is the number of arms in (the \emph{order} of) the star of \textbf{k}, 
then the induced irrep $\bfit{D}^{\mathbf{*k},\,i}(\gro{G})$ has the dimension $r\times s$ and
its matrices can be written in the form: 
\begin{equation}\label{sgirrep}
\bfit{D}^{\mathbf{*k},\,i}(\{R|\,\mathbf{v}\}_{mp,nq})=
\bfit{M}(\{R|\,\mathbf{v}\}_{m,n})\bfit{D}^{\mathbf{k},\,i}
(\{R^{\mathbf{k}}|\,\mathbf{v}^{\mathbf{k}}\}_{p,q}) \,,
\end{equation}
where $\{R^{\mathbf{k}}|\,\mathbf{v}^{\mathbf{k}}\}=
(\grel{q}_m)^{-1}\,\{R|\,\mathbf{v}\}\,\grel{q}_n$
is an element of the little group $\gro{G}^{\mathbf{k}}$, with $n,m=1,\ldots,s$. Because the $s \times s$ \emph{induction matrix} $\bfit{M}(\{R|\,\mathbf{v}\})$ is a monomial matrix,  the matrices 
$\bfit{D}^{\mathbf{*k},\,i}(\{R|\,\mathbf{v}\})$ have a block structure with exactly one 
non-zero $(r \times r)$ block in every column and every row; the block is the 
matrix $\bfit{D}^{\mathbf{k},\,i}(\{R^{\mathbf{k}}|\,\mathbf{v}^{\mathbf{k}}\})$, and 
$\{R^{\mathbf{k}}|\,\mathbf{v}^{\mathbf{k}}\}$ is fixed by the condition $\{R^{\mathbf{k}}|\,\mathbf{v}^{\mathbf{k}}\}=(\grel{q}_m)^{-1}\,\{R|\,\mathbf{v}\}\,\grel{q}_n\,\in\gro{G}^{\mathbf{k}}$.

\begin{acknowledgments}
The work of LE, GF and MIA was supported by the Government of the Basque Country (project IT779-13) and the Spanish Ministry of Economy and Competitiveness and FEDER funds (project MAT2015-66441-P). The work of MVG was supported by FIS2016- 75862-P and FIS2013-48286-C2-1-P national projects of the Spanish MINECO. ZW and BAB, as well as part of the development of the initial theory and further ab-initio work, were supported by the NSF EAGER Grant No. DMR-1643312, ONR - N00014-14-1-0330, ARO MURI W911NF-12-1-0461, and NSF-MRSEC DMR-1420541. The development of the practical part of the theory, tables, some of the code development, and ab-initio work was funded by Department of Energy de-sc0016239, Simons Investigator Award, the Packard Foundation, and the Schmidt Fund for Innovative Research. BB, JC, ZW, and BAB acknowledge the hospitality of the Donostia International Physics Center, where parts of this work were carried out. JC acknowledges the hospitality of the Kavli Institute for Theoretical Physics, and BAB acknowledges the hospitality and support of the \'Ecole Normale Sup\'erieure and Laboratoire de Physique Th\'eorique et Hautes Energies.
\end{acknowledgments}

\bibliography{elcoro}

\begin{thebibliography}{28}%
\makeatletter
\providecommand \@ifxundefined [1]{%
 \@ifx{#1\undefined}
}%
\providecommand \@ifnum [1]{%
 \ifnum #1\expandafter \@firstoftwo
 \else \expandafter \@secondoftwo
 \fi
}%
\providecommand \@ifx [1]{%
 \ifx #1\expandafter \@firstoftwo
 \else \expandafter \@secondoftwo
 \fi
}%
\providecommand \natexlab [1]{#1}%
\providecommand \enquote  [1]{``#1''}%
\providecommand \bibnamefont  [1]{#1}%
\providecommand \bibfnamefont [1]{#1}%
\providecommand \citenamefont [1]{#1}%
\providecommand \href@noop [0]{\@secondoftwo}%
\providecommand \href [0]{\begingroup \@sanitize@url \@href}%
\providecommand \@href[1]{\@@startlink{#1}\@@href}%
\providecommand \@@href[1]{\endgroup#1\@@endlink}%
\providecommand \@sanitize@url [0]{\catcode `\\12\catcode `\$12\catcode
  `\&12\catcode `\#12\catcode `\^12\catcode `\_12\catcode `\%12\relax}%
\providecommand \@@startlink[1]{}%
\providecommand \@@endlink[0]{}%
\providecommand \url  [0]{\begingroup\@sanitize@url \@url }%
\providecommand \@url [1]{\endgroup\@href {#1}{\urlprefix }}%
\providecommand \urlprefix  [0]{URL }%
\providecommand \Eprint [0]{\href }%
\providecommand \doibase [0]{http://dx.doi.org/}%
\providecommand \selectlanguage [0]{\@gobble}%
\providecommand \bibinfo  [0]{\@secondoftwo}%
\providecommand \bibfield  [0]{\@secondoftwo}%
\providecommand \translation [1]{[#1]}%
\providecommand \BibitemOpen [0]{}%
\providecommand \bibitemStop [0]{}%
\providecommand \bibitemNoStop [0]{.\EOS\space}%
\providecommand \EOS [0]{\spacefactor3000\relax}%
\providecommand \BibitemShut  [1]{\csname bibitem#1\endcsname}%
\let\auto@bib@innerbib\@empty
\bibitem [{\citenamefont {Aroyo}\ \emph {et~al.}(2006)\citenamefont {Aroyo},
  \citenamefont {Perez-Mato}, \citenamefont {Capillas}, \citenamefont
  {Kroumova}, \citenamefont {Ivantchev}, \citenamefont {Madariaga},
  \citenamefont {Kirov},\ and\ \citenamefont {Wondratschek}}]{aroyo2006}%
  \BibitemOpen
  \bibfield  {author} {\bibinfo {author} {\bibfnamefont {M.~I.}\ \bibnamefont
  {Aroyo}}, \bibinfo {author} {\bibfnamefont {J.~M.}\ \bibnamefont
  {Perez-Mato}}, \bibinfo {author} {\bibfnamefont {C.}~\bibnamefont
  {Capillas}}, \bibinfo {author} {\bibfnamefont {E.}~\bibnamefont {Kroumova}},
  \bibinfo {author} {\bibfnamefont {S.}~\bibnamefont {Ivantchev}}, \bibinfo
  {author} {\bibfnamefont {G.}~\bibnamefont {Madariaga}}, \bibinfo {author}
  {\bibfnamefont {A.}~\bibnamefont {Kirov}}, \ and\ \bibinfo {author}
  {\bibfnamefont {H.}~\bibnamefont {Wondratschek}},\ }\href@noop {} {\bibfield
  {journal} {\bibinfo  {journal} {Z. Kristallogr.}\ }\textbf {\bibinfo {volume}
  {221}},\ \bibinfo {pages} {15} (\bibinfo {year} {2006})}\BibitemShut
  {NoStop}%
\bibitem [{\citenamefont {Aroyo}(2016)}]{ita}%
  \BibitemOpen
  \bibfield  {author} {\bibinfo {author} {\bibfnamefont {M.~I.}\ \bibnamefont
  {Aroyo}},\ }\href@noop {} {\emph {\bibinfo {title} {International Tables for
  Crystallography, Vol. A: Space-Group Symmetry. 6th Edition}}}\ (\bibinfo
  {publisher} {Wiley},\ \bibinfo {address} {Chichester},\ \bibinfo {year}
  {2016})\BibitemShut {NoStop}%
\bibitem [{\citenamefont {Wondratschek}\ and\ \citenamefont
  {Muller}(2011)}]{ita1}%
  \BibitemOpen
  \bibfield  {author} {\bibinfo {author} {\bibfnamefont {H.}~\bibnamefont
  {Wondratschek}}\ and\ \bibinfo {author} {\bibfnamefont {U.}~\bibnamefont
  {Muller}},\ }\href@noop {} {\emph {\bibinfo {title} {International Tables for
  Crystallography, Vol. A1: Symmetry relations between space groups.}}}\
  (\bibinfo  {publisher} {Wiley},\ \bibinfo {address} {Chichester},\ \bibinfo
  {year} {2011})\BibitemShut {NoStop}%
\bibitem [{\citenamefont {Kopsky}\ and\ \citenamefont {Litvin}(2010)}]{ite}%
  \BibitemOpen
  \bibfield  {author} {\bibinfo {author} {\bibfnamefont {V.}~\bibnamefont
  {Kopsky}}\ and\ \bibinfo {author} {\bibfnamefont {D.~B.}\ \bibnamefont
  {Litvin}},\ }\href@noop {} {\emph {\bibinfo {title} {International Tables for
  Crystallography, Vol. E: Subperiodic groups}}}\ (\bibinfo  {publisher}
  {Wiley},\ \bibinfo {address} {Chichester},\ \bibinfo {year}
  {2010})\BibitemShut {NoStop}%
\bibitem [{\citenamefont {Perez-Mato}\ \emph {et~al.}(2015)\citenamefont
  {Perez-Mato}, \citenamefont {Gallego}, \citenamefont {Tasci}, \citenamefont
  {Elcoro}, \citenamefont {de~la Flor},\ and\ \citenamefont
  {Aroyo}}]{perezmato2015}%
  \BibitemOpen
  \bibfield  {author} {\bibinfo {author} {\bibfnamefont {J.~M.}\ \bibnamefont
  {Perez-Mato}}, \bibinfo {author} {\bibfnamefont {S.~V.}\ \bibnamefont
  {Gallego}}, \bibinfo {author} {\bibfnamefont {E.~S.}\ \bibnamefont {Tasci}},
  \bibinfo {author} {\bibfnamefont {L.}~\bibnamefont {Elcoro}}, \bibinfo
  {author} {\bibfnamefont {G.}~\bibnamefont {de~la Flor}}, \ and\ \bibinfo
  {author} {\bibfnamefont {M.~I.}\ \bibnamefont {Aroyo}},\ }\href@noop {}
  {\bibfield  {journal} {\bibinfo  {journal} {Annu. Rev. Mater. Res.}\ }\textbf
  {\bibinfo {volume} {45}},\ \bibinfo {pages} {217} (\bibinfo {year}
  {2015})}\BibitemShut {NoStop}%
\bibitem [{\citenamefont {Bethe}(1929)}]{bethe1929}%
  \BibitemOpen
  \bibfield  {author} {\bibinfo {author} {\bibfnamefont {H.~A.}\ \bibnamefont
  {Bethe}},\ }\href@noop {} {\bibfield  {journal} {\bibinfo  {journal} {Annalen
  der Physik}\ }\textbf {\bibinfo {volume} {3}},\ \bibinfo {pages} {133}
  (\bibinfo {year} {1929})}\BibitemShut {NoStop}%
\bibitem [{\citenamefont {Opechowski}(1940)}]{opechowski1940}%
  \BibitemOpen
  \bibfield  {author} {\bibinfo {author} {\bibfnamefont {W.}~\bibnamefont
  {Opechowski}},\ }\href@noop {} {\bibfield  {journal} {\bibinfo  {journal}
  {Physica}\ }\textbf {\bibinfo {volume} {7}},\ \bibinfo {pages} {552}
  (\bibinfo {year} {1940})}\BibitemShut {NoStop}%
\bibitem [{\citenamefont {Bradley}\ and\ \citenamefont
  {Cracknell}(1972)}]{bradley1972}%
  \BibitemOpen
  \bibfield  {author} {\bibinfo {author} {\bibfnamefont {C.~J.}\ \bibnamefont
  {Bradley}}\ and\ \bibinfo {author} {\bibfnamefont {A.~P.}\ \bibnamefont
  {Cracknell}},\ }\href@noop {} {\emph {\bibinfo {title} {The Mathematical
  Theory of Symmetry in Solids}}}\ (\bibinfo  {publisher} {Clarendon Press},\
  \bibinfo {address} {Oxford},\ \bibinfo {year} {1972})\BibitemShut {NoStop}%
\bibitem [{\citenamefont {Altmann}\ and\ \citenamefont
  {Herzig}(1994)}]{altmann1994}%
  \BibitemOpen
  \bibfield  {author} {\bibinfo {author} {\bibfnamefont {S.~L.}\ \bibnamefont
  {Altmann}}\ and\ \bibinfo {author} {\bibfnamefont {P.}~\bibnamefont
  {Herzig}},\ }\href@noop {} {\emph {\bibinfo {title} {Point Group Theory
  Tables}}}\ (\bibinfo  {publisher} {Clarendon Press},\ \bibinfo {address}
  {Oxford},\ \bibinfo {year} {1994})\BibitemShut {NoStop}%
\bibitem [{\citenamefont {Cracknell}\ \emph {et~al.}(1979)\citenamefont
  {Cracknell}, \citenamefont {Davies}, \citenamefont {Miller},\ and\
  \citenamefont {Love}}]{cracknell1979}%
  \BibitemOpen
  \bibfield  {author} {\bibinfo {author} {\bibfnamefont {A.~P.}\ \bibnamefont
  {Cracknell}}, \bibinfo {author} {\bibfnamefont {B.~L.}\ \bibnamefont
  {Davies}}, \bibinfo {author} {\bibfnamefont {S.~C.}\ \bibnamefont {Miller}},
  \ and\ \bibinfo {author} {\bibfnamefont {W.~F.}\ \bibnamefont {Love}},\
  }\href@noop {} {\emph {\bibinfo {title} {Kronecker Product Tables, 1, General
  introduction and Tables of Irreducible Representations of Space groups}}}\
  (\bibinfo  {publisher} {IFI, Plenum},\ \bibinfo {address} {New York},\
  \bibinfo {year} {1979})\BibitemShut {NoStop}%
\bibitem [{\citenamefont {Glazer}\ \emph {et~al.}(2014)\citenamefont {Glazer},
  \citenamefont {Aroyo},\ and\ \citenamefont {Authier}}]{glazer2014}%
  \BibitemOpen
  \bibfield  {author} {\bibinfo {author} {\bibfnamefont {A.~M.}\ \bibnamefont
  {Glazer}}, \bibinfo {author} {\bibfnamefont {M.~I.}\ \bibnamefont {Aroyo}}, \
  and\ \bibinfo {author} {\bibfnamefont {A.}~\bibnamefont {Authier}},\
  }\href@noop {} {\bibfield  {journal} {\bibinfo  {journal} {Acta Crystallogr.
  A}\ }\textbf {\bibinfo {volume} {70}},\ \bibinfo {pages} {300} (\bibinfo
  {year} {2014})}\BibitemShut {NoStop}%
\bibitem [{\citenamefont {Stokes}\ \emph {et~al.}(2013)\citenamefont {Stokes},
  \citenamefont {Campbell},\ and\ \citenamefont {Cordes}}]{stokes2013}%
  \BibitemOpen
  \bibfield  {author} {\bibinfo {author} {\bibfnamefont {H.~T.}\ \bibnamefont
  {Stokes}}, \bibinfo {author} {\bibfnamefont {B.~J.}\ \bibnamefont
  {Campbell}}, \ and\ \bibinfo {author} {\bibfnamefont {R.}~\bibnamefont
  {Cordes}},\ }\href@noop {} {\bibfield  {journal} {\bibinfo  {journal} {Acta
  Cryst.}\ }\textbf {\bibinfo {volume} {A69}},\ \bibinfo {pages} {388}
  (\bibinfo {year} {2013})}\BibitemShut {NoStop}%
\bibitem [{\citenamefont {Herring}(1937)}]{herring1937}%
  \BibitemOpen
  \bibfield  {author} {\bibinfo {author} {\bibfnamefont {C.}~\bibnamefont
  {Herring}},\ }\href@noop {} {\bibfield  {journal} {\bibinfo  {journal} {Phys.
  Rev.}\ }\textbf {\bibinfo {volume} {52}},\ \bibinfo {pages} {361} (\bibinfo
  {year} {1937})}\BibitemShut {NoStop}%
\bibitem [{\citenamefont {Koster}\ \emph {et~al.}(1963)\citenamefont {Koster},
  \citenamefont {Dimmock}, \citenamefont {Wheeler},\ and\ \citenamefont
  {Statz}}]{koster1963}%
  \BibitemOpen
  \bibfield  {author} {\bibinfo {author} {\bibfnamefont {G.~F.}\ \bibnamefont
  {Koster}}, \bibinfo {author} {\bibfnamefont {J.~O.}\ \bibnamefont {Dimmock}},
  \bibinfo {author} {\bibfnamefont {R.~G.}\ \bibnamefont {Wheeler}}, \ and\
  \bibinfo {author} {\bibfnamefont {H.}~\bibnamefont {Statz}},\ }\href@noop {}
  {\emph {\bibinfo {title} {Properties of the Thirty-Two Point groups}}}\
  (\bibinfo  {publisher} {MIT Press},\ \bibinfo {address} {Cambridge, MA},\
  \bibinfo {year} {1963})\BibitemShut {NoStop}%
\bibitem [{\citenamefont {Mulliken}(1933)}]{mulliken1933}%
  \BibitemOpen
  \bibfield  {author} {\bibinfo {author} {\bibfnamefont {R.~S.}\ \bibnamefont
  {Mulliken}},\ }\href@noop {} {\bibfield  {journal} {\bibinfo  {journal}
  {Phys. Rev.}\ }\textbf {\bibinfo {volume} {43}},\ \bibinfo {pages} {279}
  (\bibinfo {year} {1933})}\BibitemShut {NoStop}%
\bibitem [{\citenamefont {Miller}\ and\ \citenamefont
  {Love}(1967)}]{miller1967}%
  \BibitemOpen
  \bibfield  {author} {\bibinfo {author} {\bibfnamefont {S.~C.}\ \bibnamefont
  {Miller}}\ and\ \bibinfo {author} {\bibfnamefont {W.~F.}\ \bibnamefont
  {Love}},\ }\href@noop {} {\emph {\bibinfo {title} {Tables of irreducible
  representations of space groups and co-representations of magnetic groups}}}\
  (\bibinfo  {publisher} {Pruett},\ \bibinfo {address} {Boulder, CO},\ \bibinfo
  {year} {1967})\BibitemShut {NoStop}%
\bibitem [{\citenamefont {Dresselhaus}\ \emph {et~al.}(2008)\citenamefont
  {Dresselhaus}, \citenamefont {Dresselhaus},\ and\ \citenamefont
  {Jorio}}]{dresselhaus2008}%
  \BibitemOpen
  \bibfield  {author} {\bibinfo {author} {\bibfnamefont {M.~S.}\ \bibnamefont
  {Dresselhaus}}, \bibinfo {author} {\bibfnamefont {G.}~\bibnamefont
  {Dresselhaus}}, \ and\ \bibinfo {author} {\bibfnamefont {A.}~\bibnamefont
  {Jorio}},\ }\href@noop {} {\emph {\bibinfo {title} {Group Theory. Application
  to the Physics of Condensed Matter.}}}\ (\bibinfo  {publisher}
  {Springer-Verlag},\ \bibinfo {address} {Berlin},\ \bibinfo {year}
  {2008})\BibitemShut {NoStop}%
\bibitem [{\citenamefont {Bradlyn}\ \emph
  {et~al.}(2017{\natexlab{a}})\citenamefont {Bradlyn}, \citenamefont {Elcoro},
  \citenamefont {Cano}, \citenamefont {Vergniory}, \citenamefont {Wang},
  \citenamefont {Felser}, \citenamefont {Aroyo},\ and\ \citenamefont
  {Bernevig}}]{NaturePaper}%
  \BibitemOpen
  \bibfield  {author} {\bibinfo {author} {\bibfnamefont {B.}~\bibnamefont
  {Bradlyn}}, \bibinfo {author} {\bibfnamefont {L.}~\bibnamefont {Elcoro}},
  \bibinfo {author} {\bibfnamefont {J.}~\bibnamefont {Cano}}, \bibinfo {author}
  {\bibfnamefont {M.~G.}\ \bibnamefont {Vergniory}}, \bibinfo {author}
  {\bibfnamefont {Z.}~\bibnamefont {Wang}}, \bibinfo {author} {\bibfnamefont
  {C.}~\bibnamefont {Felser}}, \bibinfo {author} {\bibfnamefont {M.~I.}\
  \bibnamefont {Aroyo}}, \ and\ \bibinfo {author} {\bibfnamefont {B.~A.}\
  \bibnamefont {Bernevig}},\ }\href@noop {} {\bibfield  {journal} {\bibinfo
  {journal} {Nature}\ }\textbf {\bibinfo {volume} {547}},\ \bibinfo {pages}
  {298} (\bibinfo {year} {2017}{\natexlab{a}})}\BibitemShut {NoStop}%
\bibitem [{\citenamefont {Evarestov}\ and\ \citenamefont
  {Smirnov}(1997)}]{evarestov1997}%
  \BibitemOpen
  \bibfield  {author} {\bibinfo {author} {\bibfnamefont {R.~A.}\ \bibnamefont
  {Evarestov}}\ and\ \bibinfo {author} {\bibfnamefont {V.~P.}\ \bibnamefont
  {Smirnov}},\ }\href@noop {} {\emph {\bibinfo {title} {Site Symmetry in
  Crystals}}}\ (\bibinfo  {publisher} {Springer-Verlag},\ \bibinfo {address}
  {Berlin},\ \bibinfo {year} {1997})\BibitemShut {NoStop}%
\bibitem [{\citenamefont {Kitaev}\ \emph {et~al.}(1997)\citenamefont {Kitaev},
  \citenamefont {Panfilov}, \citenamefont {Tronc},\ and\ \citenamefont
  {Evarestov}}]{kitaev1997}%
  \BibitemOpen
  \bibfield  {author} {\bibinfo {author} {\bibfnamefont {Y.~E.}\ \bibnamefont
  {Kitaev}}, \bibinfo {author} {\bibfnamefont {A.~G.}\ \bibnamefont
  {Panfilov}}, \bibinfo {author} {\bibfnamefont {P.}~\bibnamefont {Tronc}}, \
  and\ \bibinfo {author} {\bibfnamefont {R.~A.}\ \bibnamefont {Evarestov}},\
  }\href@noop {} {\bibfield  {journal} {\bibinfo  {journal} {J. Phys.: Condens.
  Matter}\ }\textbf {\bibinfo {volume} {9}},\ \bibinfo {pages} {257} (\bibinfo
  {year} {1997})}\BibitemShut {NoStop}%
\bibitem [{\citenamefont {Zak}(1982)}]{zak1982}%
  \BibitemOpen
  \bibfield  {author} {\bibinfo {author} {\bibfnamefont {J.}~\bibnamefont
  {Zak}},\ }\href@noop {} {\bibfield  {journal} {\bibinfo  {journal} {Phys.
  Rev. B}\ }\textbf {\bibinfo {volume} {26}},\ \bibinfo {pages} {3010}
  (\bibinfo {year} {1982})}\BibitemShut {NoStop}%
\bibitem [{\citenamefont {Michel}\ and\ \citenamefont
  {Zak}(1992)}]{michel1992}%
  \BibitemOpen
  \bibfield  {author} {\bibinfo {author} {\bibfnamefont {L.}~\bibnamefont
  {Michel}}\ and\ \bibinfo {author} {\bibfnamefont {J.}~\bibnamefont {Zak}},\
  }\href@noop {} {\bibfield  {journal} {\bibinfo  {journal} {Europhysics
  Letters}\ }\textbf {\bibinfo {volume} {18}},\ \bibinfo {pages} {239}
  (\bibinfo {year} {1992})}\BibitemShut {NoStop}%
\bibitem [{\citenamefont {Zeiner}\ \emph {et~al.}(2000)\citenamefont {Zeiner},
  \citenamefont {Dirl},\ and\ \citenamefont {Davies}}]{zeiner2000}%
  \BibitemOpen
  \bibfield  {author} {\bibinfo {author} {\bibfnamefont {P.}~\bibnamefont
  {Zeiner}}, \bibinfo {author} {\bibfnamefont {R.}~\bibnamefont {Dirl}}, \ and\
  \bibinfo {author} {\bibfnamefont {B.~L.}\ \bibnamefont {Davies}},\
  }\href@noop {} {\bibfield  {journal} {\bibinfo  {journal} {J. Phys. A: Math.
  Gen.}\ }\textbf {\bibinfo {volume} {33}},\ \bibinfo {pages} {1631} (\bibinfo
  {year} {2000})}\BibitemShut {NoStop}%
\bibitem [{\citenamefont {Cano}\ \emph {et~al.}(2017)\citenamefont {Cano},
  \citenamefont {Bradlyn}, \citenamefont {Wang}, \citenamefont {Elcoro},
  \citenamefont {Vergniory}, \citenamefont {Felser}, \citenamefont {Aroyo},\
  and\ \citenamefont {Bernevig}}]{EBRtheory}%
  \BibitemOpen
  \bibfield  {author} {\bibinfo {author} {\bibfnamefont {J.}~\bibnamefont
  {Cano}}, \bibinfo {author} {\bibfnamefont {B.}~\bibnamefont {Bradlyn}},
  \bibinfo {author} {\bibfnamefont {Z.}~\bibnamefont {Wang}}, \bibinfo {author}
  {\bibfnamefont {L.}~\bibnamefont {Elcoro}}, \bibinfo {author} {\bibfnamefont
  {M.~G.}\ \bibnamefont {Vergniory}}, \bibinfo {author} {\bibfnamefont
  {C.}~\bibnamefont {Felser}}, \bibinfo {author} {\bibfnamefont {M.~I.}\
  \bibnamefont {Aroyo}}, \ and\ \bibinfo {author} {\bibfnamefont {B.~A.}\
  \bibnamefont {Bernevig}},\ }\href@noop {} {\enquote {\bibinfo {title}
  {Building blocks of topological quantum chemistry: Elementary band
  representations},}\ } (\bibinfo {year} {2017}),\ \bibinfo {note}
  {arXiv:1709.01935 [cond-mat]}\BibitemShut {NoStop}%
\bibitem [{\citenamefont {Bacry}\ \emph {et~al.}(1988)\citenamefont {Bacry},
  \citenamefont {Michel},\ and\ \citenamefont {Zak}}]{bacry1988}%
  \BibitemOpen
  \bibfield  {author} {\bibinfo {author} {\bibfnamefont {H.}~\bibnamefont
  {Bacry}}, \bibinfo {author} {\bibfnamefont {L.}~\bibnamefont {Michel}}, \
  and\ \bibinfo {author} {\bibfnamefont {J.}~\bibnamefont {Zak}},\ }\enquote
  {\bibinfo {title} {Symmetry and classification of energy bands in
  crystals},}\ in\ \href@noop {} {\emph {\bibinfo {booktitle} {Group
  theoretical methods in Physics: Proceedings of the XVI International
  Colloquium Held at Varna, Bulgaria, June 15--20 1987}}}\ (\bibinfo
  {publisher} {Springer Berlin Heidelberg},\ \bibinfo {year} {1988})\ p.\
  \bibinfo {pages} {289}\BibitemShut {NoStop}%
\bibitem [{\citenamefont {Vergniory}\ \emph {et~al.}(2017)\citenamefont
  {Vergniory}, \citenamefont {Elcoro}, \citenamefont {Cano}, \citenamefont
  {Wang}, \citenamefont {Felser}, \citenamefont {Aroyo}, \citenamefont
  {Bernevig},\ and\ \citenamefont {Bradlyn}}]{GraphDataPaper}%
  \BibitemOpen
  \bibfield  {author} {\bibinfo {author} {\bibfnamefont {M.~G.}\ \bibnamefont
  {Vergniory}}, \bibinfo {author} {\bibfnamefont {L.}~\bibnamefont {Elcoro}},
  \bibinfo {author} {\bibfnamefont {J.}~\bibnamefont {Cano}}, \bibinfo {author}
  {\bibfnamefont {Z.}~\bibnamefont {Wang}}, \bibinfo {author} {\bibfnamefont
  {C.}~\bibnamefont {Felser}}, \bibinfo {author} {\bibfnamefont {M.~I.}\
  \bibnamefont {Aroyo}}, \bibinfo {author} {\bibfnamefont {B.~A.}\ \bibnamefont
  {Bernevig}}, \ and\ \bibinfo {author} {\bibfnamefont {B.}~\bibnamefont
  {Bradlyn}},\ }\href@noop {} {\bibfield  {journal} {\bibinfo  {journal} {Phys.
  Rev. E}\ }\textbf {\bibinfo {volume} {96}},\ \bibinfo {pages} {023310}
  (\bibinfo {year} {2017})}\BibitemShut {NoStop}%
\bibitem [{\citenamefont {Bradlyn}\ \emph
  {et~al.}(2017{\natexlab{b}})\citenamefont {Bradlyn}, \citenamefont {Elcoro},
  \citenamefont {Vergniory}, \citenamefont {Cano}, \citenamefont {Wang},
  \citenamefont {Felser}, \citenamefont {Aroyo},\ and\ \citenamefont
  {Bernevig}}]{GraphTheoryPaper}%
  \BibitemOpen
  \bibfield  {author} {\bibinfo {author} {\bibfnamefont {B.}~\bibnamefont
  {Bradlyn}}, \bibinfo {author} {\bibfnamefont {L.}~\bibnamefont {Elcoro}},
  \bibinfo {author} {\bibfnamefont {M.~G.}\ \bibnamefont {Vergniory}}, \bibinfo
  {author} {\bibfnamefont {J.}~\bibnamefont {Cano}}, \bibinfo {author}
  {\bibfnamefont {Z.}~\bibnamefont {Wang}}, \bibinfo {author} {\bibfnamefont
  {C.}~\bibnamefont {Felser}}, \bibinfo {author} {\bibfnamefont {M.~I.}\
  \bibnamefont {Aroyo}}, \ and\ \bibinfo {author} {\bibfnamefont {B.~A.}\
  \bibnamefont {Bernevig}},\ }\href@noop {} {\enquote {\bibinfo {title} {Band
  connectivity for topological quantum chemistry: Band structures as a graph
  theory problem},}\ } (\bibinfo {year} {2017}{\natexlab{b}}),\ \bibinfo {note}
  {arXiv:1709.01937 [cond-mat]}\BibitemShut {NoStop}%
\bibitem [{\citenamefont {Zak}(1960)}]{zak1960}%
  \BibitemOpen
  \bibfield  {author} {\bibinfo {author} {\bibfnamefont {J.}~\bibnamefont
  {Zak}},\ }\href@noop {} {\bibfield  {journal} {\bibinfo  {journal} {J. Math.
  Phys.}\ }\textbf {\bibinfo {volume} {1}},\ \bibinfo {pages} {165} (\bibinfo
  {year} {1960})}\BibitemShut {NoStop}%
\end{thebibliography}%
\newpage

\end{document}